\documentclass[twoside,twocolumn,reprint
 amsmath,amssymb,
 aps,
pra,
floatfix
]{revtex4-1}

\usepackage{graphicx}
\usepackage{bm}
\usepackage{braket}

\usepackage{amsmath}
\usepackage{gensymb}		
\usepackage{bm}				
\usepackage{adjustbox}		

\begin{document}

\preprint{APS/123-QED}

\title{Orbital-angular-momentum free-space optical communication via the azimuthal phase-shift}

\author{Raymond Lopez-Rios$^{1,2}$}
\author{Usman A. Javid$^{1,2}$}
\author{Qiang Lin$^{1,2,3}$}%
\email{qiang.lin@rochester.edu}
\affiliation{$^{1}$Institute Of Optics, University of Rochester, Rochester NY 14627, USA}
\affiliation{$^{2}$Center for Coherence and Quantum Optics, University of Rochester, Rochester NY 14627, USA}
\affiliation{$^{3}$Department of Electrical and Computer Engineering, University of Rochester, Rochester NY 14627, USA}

\date{\today}

\begin{abstract}
	Free-space optical communication using the orbital angular momentum (OAM) of light has garnered significant interest lately due to the potentially vast bandwidth intrinsic to the infinite Hilbert space of OAM modes. Unfortunately, OAM light beams suffer from serious distortions due to atmospheric turbulence (AT) that has become a dominant factor limiting the advance of OAM-based free-space communication. Here we propose and demonstrate a free-space communication scheme---using OAM beams and their azimuthal-mode phase-shift for keying (OAM-APSK)---which is resilient to AT-induced distortions. Combined with a digital holographic mode sorting (DHMS) technique, the proposed approach is able to achieve high signal-to-noise ratios and to maintain low modal crosstalk, even for extremely strong turbulence conditions, with magnitudes significantly beyond existing AT mitigation methods. The demonstrated OAM-APSK and DHMS schemes may now open up a great avenue for OAM-based free-space optical communication that could elegantly resolve the long-standing challenge imposed by atmospheric turbulence.
\end{abstract}
	
	\maketitle
	\twocolumngrid
	\section{Introduction}
	
	With an ever-increasing demand for greater bandwidths and speeds in wireless information transfer, free-space optical communication offers a promising solution with superior bandwidth \cite{borah2012review, khalighi2014survey}, where information can be encoded in various optical degrees of freedom such as time, wavelength, polarization, phase, and space. From the latter, orbital angular momentum (OAM) \cite{yao2011orbital}, which denotes helical wavefronts with discrete and orthogonal azimuthal modes, can function as an excellent information carrier for free-space optical communication \cite{gibson2004free, molina2007twisted}. With a potentially infinite Hilbert space for spatial-mode division multiplexing, OAM-based free-space communication has attracted significant interest in recent years for both classical and quantum-encrypted information processing \cite{willner2015optical, willner2017recent, erhard2018twisted}. 
	
	Although significant advances have been made in OAM-based communication, a major obstacle is that OAM beams are very fragile under perturbations to their complex wavefronts. One typical effect is atmospheric turbulence (AT), which has been shown to be a dominant factor in limiting the distance and capacity of OAM-based free-space communication \cite{tyler2009influence, rodenburg2012influence, ren2013atmospheric}. In the past decade, significant efforts have been devoted to mitigate the impact of AT \cite{malik2012influence, ren2014adaptive, krenn2014communication, li2014evaluation, ren2014adaptive2, ren2016atmospheric, lavery2017free, li2018atmospheric, li2018joint, yan2018controlling, liu2019single}. Current approaches are primarily based upon either adaptive optics to correct/compensate wavefront error or digital signal processing to reduce channel crosstalk \cite{willner2015optical, li2018atmospheric}, both of which unfortunately exhibit limited performance. To date, AT-induced degradation remains a critical challenge for OAM-based free-space communication. 
	
	Here we propose and demonstrate a combined encoding and decoding solution to address the AT problem. Instead of using the OAM modes $\ell$ solely as information carriers \cite{willner2015optical}, we propose using the OAM azimuthal mode phase-shift $\Delta\ell$ for keying, which is robust to AT-induced phase perturbations. Further, to bypass the effect of AT-induced intensity fluctuations, we propose a digital holographic approach for measuring $\Delta\ell$ from only spatial phase information. We show, both numerically and experimentally, that the proposed approach is able to significantly improve signal-to-noise ratios and suppress OAM channel crosstalk. By resolving the challenge imposed by atmospheric turbulence, this approach may enable more useful and interesting schemes for OAM free-space optical communication.

	\section{Concept}
	
	\subsection{OAM azimuthal phase-shift keying (OAM-APSK)}

	\begin{figure*}
		\centering
		\includegraphics[width=1\linewidth]{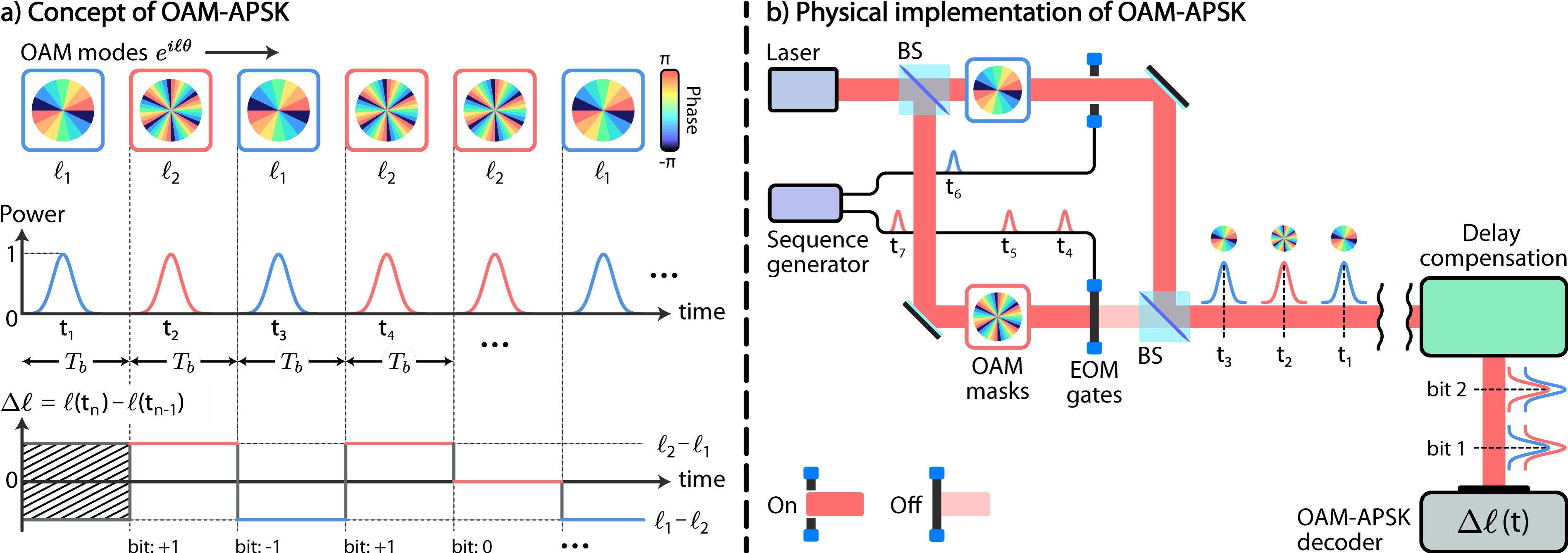}
		\caption[]{Conceptual illustration of the proposed OAM azimuthal phase-shift keying (OAM-APSK). a) The information is encoded as the azimuthal mode difference, $\Delta\ell = \ell(t_n) - \ell(t_{n-1})$, of consecutive pulses separated by the slot time $T_b$; different values of $\Delta\ell$ count as unique bits (0, +1, -1). b) Physical implementation of the OAM-APSK encoding, used only to show the fundamental operation principle. A more practical example, which can utilize the full power of the laser source, is provided in Appendix Section IV. EOM gate: electro-optic modulation gate; BS: beam splitter.}
		\label{fig:1}
	\end{figure*}
	
	The major problem of AT lies in the spatiotemporally-dependent refractive index fluctuations introduced by turbulent eddies along the path where an OAM beam propagates, imposing random phase fluctuations on OAM beam wavefronts. Our proposed approach recognizes and leverages the fact that AT fluctuates on a time scale ($\sim$ milliseconds \cite{andrews2005laser}) much longer than the slot time $T_b$ of a communication bit (\emph{e.g.}, $T_b=$ 100~ps for a bit rate of 10~Gb/s). As a result, optical pulses in adjacent time slots would experience exactly the same phase fluctuations when they propagate through AT. We thus propose an encoding scheme for OAM-based free-space communication, whose operational principle is schematically shown in Fig.~\ref{fig:1}(a).
	
	The optical pulses in each time slot exhibit the same amplitude profile $A_0(t-nT_b)$ but carry a certain OAM azimuthal charge $\ell(n)$ (where $n$ is an integer denoting the time slot number): $E_n(t,\theta) = A_0(t-nT_b)\exp[i\ell(n) \theta]$. Here $\ell(n)$ exhibits two values, $\ell(n) =$ $\ell_1$ or $\ell_2$, and $\theta$ is the azimuthal angle. Information is encoded in the OAM azimuthal charge difference, $\Delta \ell(n) = \ell(n) - \ell({n-1})$, between adjacent optical pulses, where $\Delta \ell(n) = 0$ represents bit "0" and $\Delta \ell(n) = \pm(\ell_2 - \ell_1)$ represents bit "$\pm$1" (or $|\Delta \ell(n)| = 1$ denotes bit "1"). These encoded bits are clearly shown via the two-pulse interference: 
	\begin{eqnarray}
	&& \left|E_n(t,\theta) + E_{n-1}(t, \theta)\right|^2 \nonumber\\
	&& = 2 |A_0(t-nT_b)|^2 \left(1+ \cos\{[\ell(n) - \ell({n-1})]\theta\}\right)
	\label{En_Inter}.
	\end{eqnarray}
	
	Since bits are encoded in the azimuthal phase difference between adjacent pulses, we term this encoding scheme as azimuthal phase-shift keying (APSK). APSK is similar to the differential phase-shift keying (DPSK) scheme used in fiber-optic communication \cite{agrawal2012fiber}, but the information is carried in the relative OAM degree of freedom rather than in the relative phase of pulses.
	
	When optical pulses propagate through AT, the latter imposes a random phase front $\exp[i \phi_{\rm AT}(t,x,y)]$ and the optical field becomes $E_n(t,\theta) \approx A_0(t-nT_b)\exp[i\ell(n) \theta] \exp[i \phi_{\rm AT}(t,x,y)]$. Since AT-induced fluctuations remain nearly static within a small time period $T_b$, $\phi_{\rm AT}(t,x,y) \approx \phi_{\rm AT}(t-T_b,x,y)$ for any adjacent optical pulses, and Eq.~(\ref{En_Inter}) remains intact. Clearly APSK is intrinsically immune to AT-induced phase perturbations. 
	
	In practice, OAM-APSK can be realized via a simple setup such as that shown in Fig.~\ref{fig:1}(b), where a laser is equally split into two paths, each of which is equipped with 1) a phase mask for imposing an OAM wavefront and 2) an amplitude modulator functioning as a gate. Within each time slot, only one gate is open to produce an optical pulse with a desired OAM. By using a digital communication signal to drive the electro-optic gates in each path, we will be able to produce a sequence of APSK optical pulses as shown in Fig.~\ref{fig:1}(a). This approach supports high-speed information encoding, whose speed depends only on the electro-optic gating.
	
	\subsection{Digital holographic mode sorting (DHMS)}
	\begin{figure*}
		\centering
		\includegraphics[width=1\linewidth]{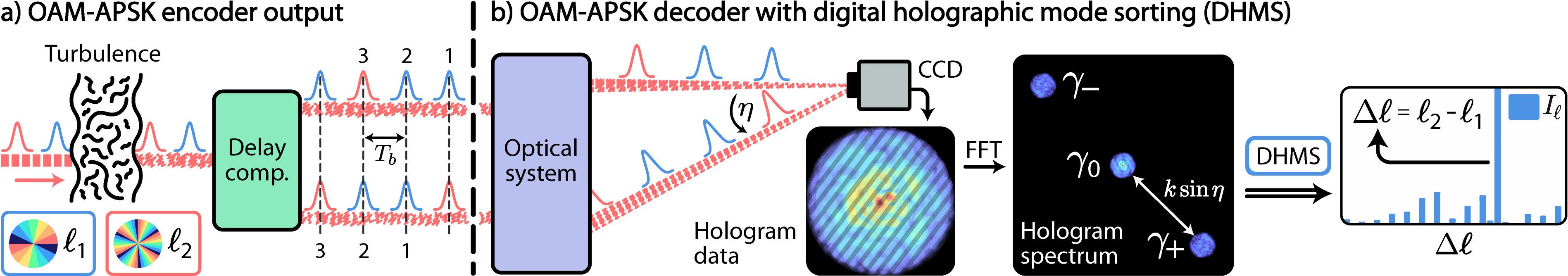}
		\caption[]{Conceptual illustration of the proposed digital holographic mode sorting (DHMS) method for decoding an OAM-APSK signal. a) Atmospheric turbulence distorts the spatial information encoded in OAM modes $\ell_1$ and $\ell_2$ before the delay compensation. b) Delayed beams interfere at the detector, which records intensity as a digital hologram due to an initial tilt $\eta$ in one beam. We then take a fast Fourier transform (FFT) of the data and use the spectrum for sorting through OAM modes (DHMS) for values of $\Delta\ell = \ell_2 - \ell_1$.}
		\label{fig:2}
	\end{figure*}
	
	\begin{figure*}
		\centering
		\includegraphics[width=\linewidth]{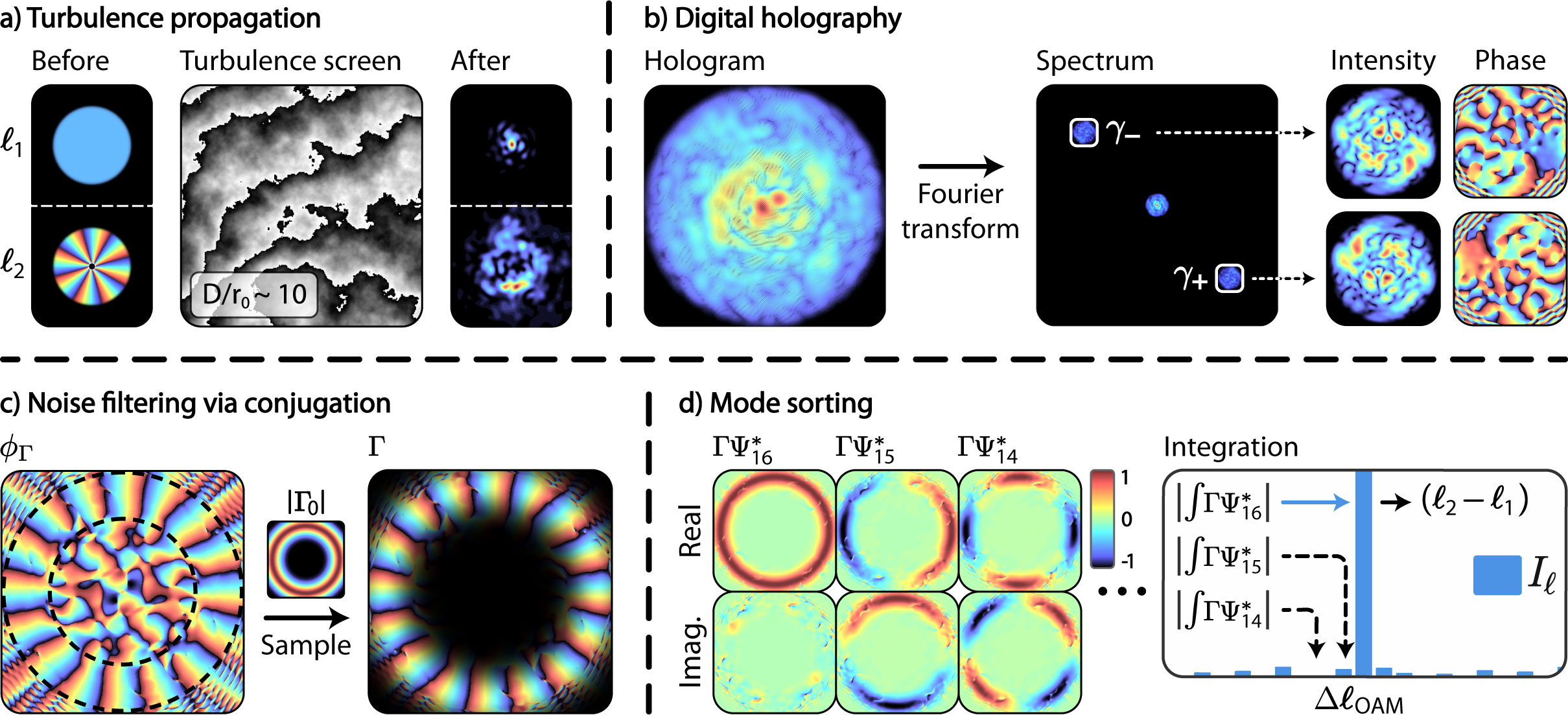}
		\caption[]{Example of a numerical simulation showing the effect of DHMS. a) Phase distributions of two OAM beams with modes $\ell_1 = 0$ and $\ell_2 = 8$, and their intensity profiles after propagating through simulated AT  (strength factor $D/r_0 \sim 10$). b) Holographic interference pattern at the camera, its angular spectrum, and the amplitude/phase of the spectral fields $\gamma_+$ and $\gamma_-$. c) The phase $\phi_{\Gamma}$ of $\gamma_+ \gamma_-^*$ and the filtered field $\Gamma \equiv \Gamma_0 \exp(i\phi_{\Gamma})$ after sampling with the filter function $\Gamma_0$. The two dashed curves in the figure of $\phi_\Gamma$ are used to indicate the annular section with negligible AT perturbations. d) Real/imaginary parts of the products between $\Gamma$ and several OAM basis functions $\Psi_{\ell} = \exp(i\ell\theta)$, and the results of mode sorting via integration, where $I_{\ell} = \left| \int \Gamma \Psi_{\ell}^* \right|$. DHMS yields a peak for the $\Psi_{\ell}^*$ that corresponds to twice the mode difference $\Delta\ell$, i.e.~here  $I_{2\Delta\ell} = I_{16}$ yields a peak at the APSK signal $\Delta\ell = \ell_2 - \ell_1 = 8$.}
		\label{fig:3}
	\end{figure*}

	The APSK scheme described above is able to address the impact of phase fluctuations. However, for sufficiently long propagation distance or strong enough turbulence, phase fluctuations will be transferred to intensity fluctuations, a phenomenon also known as scintillation \cite{johansson1994simulation, mabena2019optical}, which cannot be solved by the APSK interference itself. To resolve this issue, we propose a digital holographic approach for sorting (DHMS) and decoding the OAM-APSK signal. Digital holography is a well-known technique used in imaging applications for obtaining a faithful measurement of the phase or complex amplitude of a scattered light-field \cite{mann2005high, kemper2008digital, merola2017tomographic}. Recently, it has been used for improving free-space imaging across turbulent media \cite{marron2009atmospheric, kim2012adaptive, banet2018digital}, and these developments inspired us to use digital holography in OAM-APSK for AT compensation.
	
	Figure~\ref{fig:2} shows the proposed holographic decoding scheme. At the receiver end of an OAM-APSK link the signal beam is equally split into two paths, where one path is delayed by the slot time $T_b$. The two beams are then projected onto a camera which records their interference pattern, where one beam is tilted at an angle $\eta$ with respect to the other. As a result, the angular spectrum of the recorded interference pattern will be distributed over three different regions: one around zero-frequency with a spectral amplitude of $\gamma_0(k_x,k_y)$, and the other two around spatial frequencies of $\pm k\sin\eta$ with spectral amplitudes of $\gamma_{\pm}(k_x,k_y)$, where $k_x$ and $k_y$ are the spatial frequency coordinates. Since the recorded OAM interference carries the APSK mode $\Delta\ell$, both non-zero frequency spectral regions $\gamma_{\pm}$ also carry the modes $\pm\Delta\ell$ (one is the sign-inverted complex conjugate of the other). Consequently, the product $\gamma_+\gamma_-^*$ carries the APSK signal $2 \Delta \ell \theta$. Interestingly, it turns out that AT-induced fluctuations are dramatically suppressed in the \textit{phase} of $\gamma_{+}\gamma_{-}^*$,
	\begin{equation}
	\phi_{\Gamma} = \arg\{\gamma_+\gamma_-^*\}.
	\end{equation}
	For obtaining this final result, we provide a detailed theoretical walk-through in Appendix Section I.
	
	\section{Simulation}
	
	To show the effect of the proposed APSK-DHMS, we performed detailed numerical simulations. For simulating AT, we used the modified von-Karman power spectrum \cite{lutomirski1971wave}:
	\begin{equation}
	\widetilde{\phi}_{\mathrm{AT}}(k) = 0.49 r_0 ^{-5/3} \frac{e^{-k^2/k_m^2}}{(k^2 + k_0^2)^{11/6}}, \label{Phi}
	\end{equation}
	where $k$ is the spatial frequency magnitude $k = (k_x^2 + k_y^2)^{1/2}$, $r_0$ is the Fried coherence diameter, and $k_m$ and $k_0$ represent the inner and outer scale frequency, respectively (the latter two are related to the inner and outer scale constants $l_0$ and $L_0$ as $k_m = 5.92 \times 2\pi/l_0$ and $k_0 = 2\pi/L_0$). To cover a wide noise bandwidth, we set $l_0$ and $L_0$ respectively as 0.01 and 100 times $L$, where $L$ is 5 cm. Using the Fourier transform method, $\widetilde{\phi}_{\rm AT}$ is used to produce the AT phase $\phi_{\rm AT}(x,y)$ \cite{mcglamery1976computer}, and in all simulations we implement $\phi_{\mathrm{AT}}$ as a single thin phase screen. More details on both our AT model and the simulation flow are given in Appendix Section II. 
	
	Figure~\ref{fig:3} shows an example simulation where two beams carry OAM modes of $\ell_1 =0$ and $\ell_2 =8$. After propagation through AT, both field amplitudes are significantly distorted [Fig.~\ref{fig:3}(a)]. As a result, the recorded digital hologram shows a noisy interference pattern, from which it is difficult to extract the OAM charge difference between the two beams [Fig.~\ref{fig:3}(b)]. Since one beam was tilted at an angle of 0.186$^\mathrm{\circ}$ (determined by the wavelength and grid dimensions), the hologram spectrum clearly shows three separate regions, of which $\gamma_+$ and $\gamma_-$ still exhibit noise in their amplitude and phase. However, as shown in Fig.~\ref{fig:3}(c), $\phi_{\Gamma}$ has a clean azimuthal pattern around an annular section with negligible AT perturbations. Therefore, by filtering out this phase section (say, by multiplying with an annular profile $\Gamma_0$ shown in Fig.~\ref{fig:3}(c)), we obtain the clean field $\Gamma \equiv \Gamma_0 \exp(i\phi_{\Gamma})$, which clearly recovers the desired phase distribution of $2 \Delta \ell \theta$. To quantify the signal purity for an OAM mode $\ell$ or $\Delta\ell$, we perform \textit{conjugate mode sorting} \cite{gibson2004free} and calculate the normalized inner product $I_{\ell}$:
	\begin{equation}
	I_{\ell} = \left<\Gamma|\Psi_{\ell}^*\right> = \left| \int \Gamma \Psi_{\ell}^* \right|,
	\label{eq:dhms}
	\end{equation}
	where $\Psi_{\ell}$ denotes an OAM basis state $\exp(i \ell \theta)$. Fig.~\ref{fig:3}(d) shows the mode sorting amplitudes, where we accurately retrieve the signal $\Delta\ell = \ell_2-\ell_1$. Having obtained this result while accumulating only a negligible amount of modal crosstalk, this example simulation clearly shows the effectiveness of DHMS.
	
	\begin{figure}
		[b!]
		\centering
		\includegraphics[width=1\linewidth]{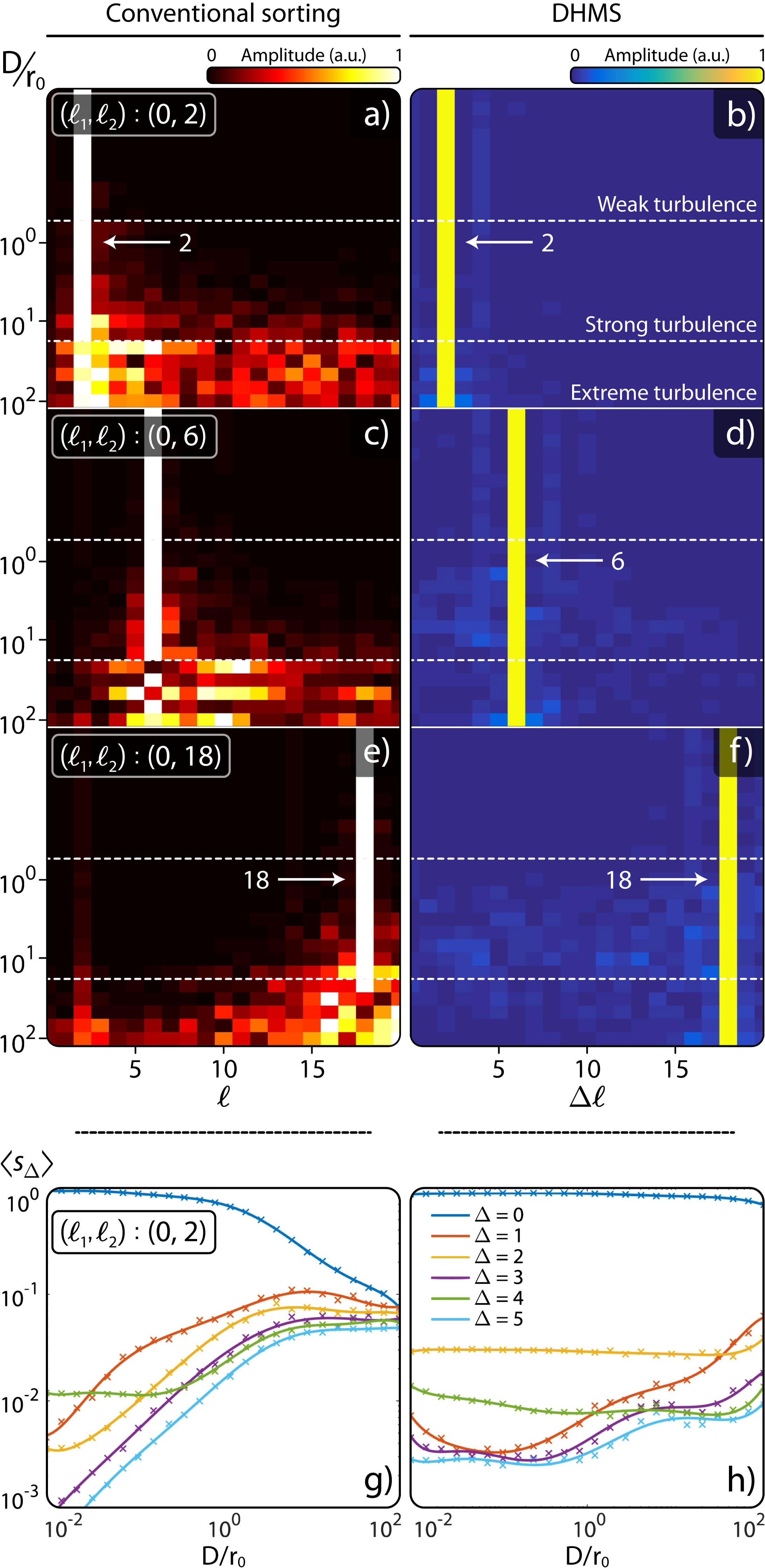}
		\caption[]{Crosstalk matrix simulations of APSK-DHMS versus the turbulence strength parameter $D/r_0$. The right column shows the result for APSK-DHMS and the left one shows that for conventional mode sorting as a comparison. The cases for three different OAM mode pairs, $(\ell_1,\ell_2) = \{(0,2),(0,6),(0,18)\}$, are shown in (a)-(b), (c)-(d), and (e)-(f), respectively. Color maps represent the amplitude (normalized for each row) of decoded OAM mode signals after either conventional sorting (red in a, c, e) or DHMS (blue in b, d, f). (g) and (h) plot the mode-specific power $\left<s_\Delta\right>$ as a function of $D/r_0$, for the case of $(\ell_1,\ell_2) = (0, 2)$. $\left<s_\Delta\right>$ was averaged over 100 DHMS simulations, and log-polynomial curves were fitted to the results for $\Delta = \{0, 1, 2, 3, 4, 5, 6\}$ (See Appendix Section II for fitting details).}
		\label{fig:4}
	\end{figure}
	
	To further emphasize the validity of APSK-DHMS, in Fig.~\ref{fig:4}(a-f) we plot the DHMS amplitudes versus different AT strengths and compare them side-by-side with conventional conjugate mode sorting \cite{gibson2004free} amplitudes. The latter is calculated using Eq.~(\ref{eq:dhms}), but $\Gamma$ is replaced by the complex electric field that carries OAM mode $\ell_2$ (right before this field interferes with the $\ell_1$ field and the DHMS hologram is recorded). Also, in order to make a direct comparison between conventional sorting modes $\ell$ and the DHMS azimuthal-phase shift $\Delta\ell$, we set $\ell_1 = 0$ in these simulations such that $\Delta\ell = \ell_2-\ell_1 = \ell_2$. In this way, three different cases of OAM mode pairs ($\ell_1, \ell_2$) are simulated in order to compare the performance of DHMS between low and high frequency azimuthal modes. For each case, AT strength is quantified by the ratio $D/r_0$ of the source aperture diameter $D$ to the Fried coherence diameter $r_0$, where $D/r_0$ varies between 0.01 and 100.\clearpage
	
	
	\begin{figure}
		[b!]
		\centering
		\includegraphics[width=1\linewidth]{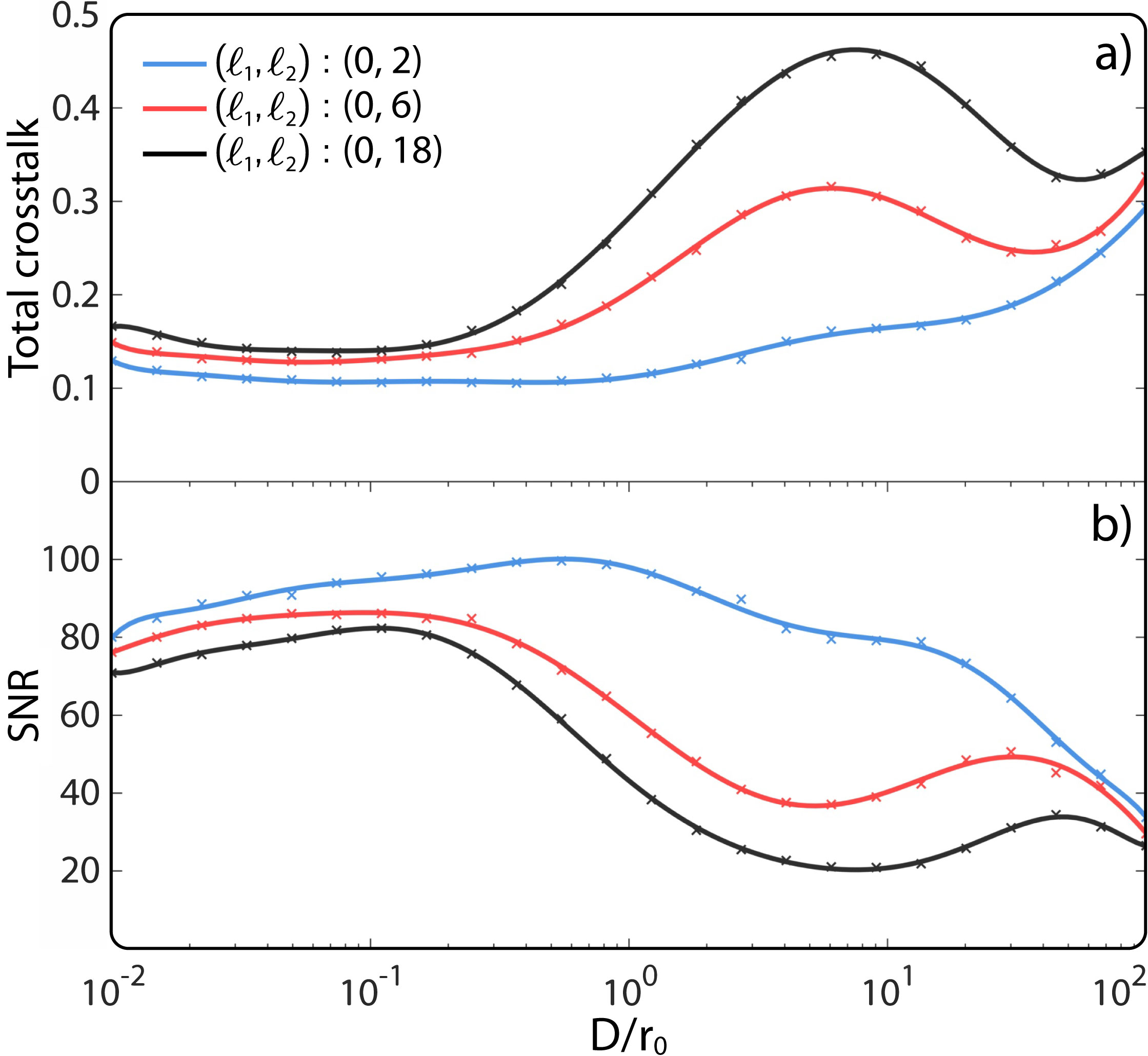}
		\caption[]{Simulation results of a) crosstalk level or percentage (XTP) and b) signal-to-noise ratio (SNR) versus $D/r_0$ for three OAM mode pairs, $(\ell_1,\ell_2) = \{(0,2), (0,6), (0,18)\}$. Log-polynomial curves were fitted to ensemble averages from 100 SNR and 100 XTP simulations (see Appendix Section II for fitting details).}
		\label{fig:5}
	\end{figure}
	
	For the (0, 2) case in Figure \ref{fig:4}(a, b), the mode signal peak appears at $\Delta\ell = \ell = 2$ in both the DHMS and conventionally sorted results. While for conventional sorting the peak begins wavering at $D/r_0 \sim 15$, in the DHMS scenario this peak does not waver at all even for $D/r_0 \sim 100$. Further, the crosstalk (or power carried by non-signal OAM modes) for the conventional sorting scenario increases starting at $D/r_0 \sim 5$ [Fig.~\ref{fig:4}(a)], but there is little crosstalk throughout the entire $D/r_0$ range in the DHMS scenario [Fig.~\ref{fig:4}(b)]. Similar results are evident for the (0, 6) and (0, 18) cases, while there is slightly more crosstalk in the DHMS results at intermediate $D/r_0$ levels $\sim 8$ [Fig.~\ref{fig:4}(d, f)]. Overall, these simulations show that the proposed APSK-DHMS combination, as opposed to conventional mode sorting, is robust to even extreme levels of turbulence and significantly minimizes crosstalk for a wide range of OAM modes.
	
	In order to take a closer look at crosstalk matrix simulations, we analyze the parameter $\left<s_{\Delta}\right>$, defined as the ensemble average (over 100 simulations) of power detected in the OAM mode $\Psi_{\ell + \Delta}$, where $\Delta$ is an integer step in the mode index $\ell$ \cite{tyler2009influence}. Specifically, $\left<s_{\Delta = 0}\right>$ denotes the power-normalized \textit{mode purity} and $\left<s_{\Delta \neq 0}\right>$ is the \textit{mode-specific crosstalk}. Figuire~\ref{fig:4}(g) and (h) plot for the case $(\ell_1, \ell_2) = (0, 2)$. As expected, $\left<s_{\Delta}\right>$ results for conventional sorting [Fig.~\ref{fig:4}(g)] are consistent with the known property of OAM modes that the power disperses homogeneously into sidebands due to increasing AT strength \cite{tyler2009influence, rodenburg2012influence}. Indeed, all $\left<s_{\Delta}\right>$ values converge to $\sim 1/21$, which is the inverse of the number of modes that are being sorted for. In strong contrast, with DHMS, the mode-specific crosstalk never increases beyond 0.1 or 10\%, and mode purity never goes below 0.69 [Fig.~\ref{fig:4}(h)] even for $D/r_0 \sim 100$. It is clearly evident that DHMS potentially has a significant advantage compared other existing AT compensation methods \cite{ren2014adaptive2, liu2019single, fu2016pre}.
	
	
	\begin{figure*}
		[t!]
		\centering
		\includegraphics[width=1\linewidth]{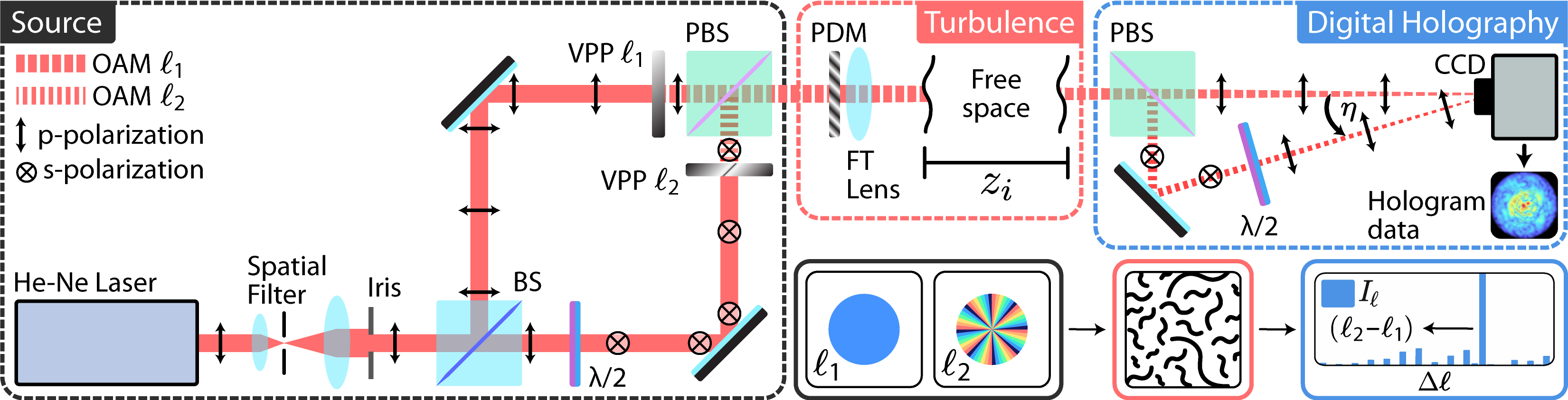}
		\caption[]{Schematic of the experimental setup. A linearly polarized 632.8-nm He-Ne laser beam is split into two arms of a Mach-Zehnder setup, where the two separated beams are imparted with OAM modes $\ell_1 = 0$ and $\ell_2 = 8$, respectively, using vortex phase plates (VPP). As discussed in the main text, for experimental convenience, we use a continuous-wave laser instead of optical pulses to prove the operation principle for transferring a single OAM APSK bit. As such, we employ polarization instead of time delay to combine and to separate the two beams. For this purpose, we rotate the polarization of one beam by 90 degrees with a half-wave ($\lambda/2$) plate. The two beams are then combined together and propagate collinearly through free space. To simulate AT, a thin phase-distorting mask (PDM) is inserted into the optical path, and a lens (FT) with a focal length of 2~m is used to simulate scaled far-field propagation along a distance $z_i \sim 2$ m. At the receiver side, the two beams are separated, one of which is normally incident onto the CCD camera. The other beam has its polarization rotated back by 90 degrees with another $\lambda/2$ plate and is then incident onto the camera with a tilt angle of $\eta$. The holographic interference pattern is recorded by the camera, which is then used to sort the OAM modes for the difference of $\ell_2$ and $\ell_1$, $\Delta \ell = \ell_2 - \ell_1$. BS: beam splitter; PBS: polarization beam splitter.}
		\label{fig:6}
	\end{figure*}

	We further quantify APSK-DHMS performance by calculating two simple metrics: the \textit{total crosstalk level or percentage} (XTP), defined as the ratio of total noise power to total power, and the associated \textit{signal-to-noise ratio} (SNR). These are given by the following expressions \cite{agrawal2012fiber}: 
	\begin{eqnarray}
	\mathrm{XTP} = \left< {\frac{ \sum_{\ell \neq \ell_2}  I_{\ell} }{\sum_{\ell}  I_{\ell} }} \right>, 
	\qquad 
	\mathrm{SNR} = \left< {\frac{ I_{\ell_2} }{\mathrm{RMS}\left\{ I_{\ell \neq \ell_2} \right\} }} \right>, \label{XTP_SNR}
	\end{eqnarray}
	where $\langle \rangle$ denotes an ensemble average over multiple instances of atmospheric turbulence and RMS$\{\cdot\}$ denotes the root mean square amplitude. Fig.~\ref{fig:5} shows both the XTP and SNR of the APSK-DHMS scheme versus AT strength for three different cases of ($\ell_1,\ell_2$). These results were calculated by averaging 100 DHMS simulations like those shown in Fig.~\ref{fig:4}(b), and in $\sum I_{\ell}$ we used a range of $\ell = \{0, 1, 2, \ldots, 20\}$.
	
	Figure~\ref{fig:5}(a) shows that XTP increases with increasing $D/r_0$, with a mostly gradual change for the (0, 2) case, but reaching a peak at $D/r_0 \sim 6$ and $\sim 7$ for cases (0, 6) and (0, 18), respectively. This trend suggests that cases with very small $\Delta\ell$, e.g.~(0, 2), yield the lowest XTP for all values of $D/r_0$, and the inverse occurs for propagations with large $\Delta\ell$, e.g.~(0, 18). The lowest value of XTP is 0.11 for (0, 2), which reaches a maximum of 0.29 at $D/r_0 \sim 100$. XTP goes to its highest value of 0.46 for (0, 18) at $D/r_0 \sim 7$, but in this case it starts off at 0.17 for $D/r_0 \sim 0.01$. 
	
	Interestingly the peaks in XTP occur for instances of AT where $D/r_0 < 10$, and this likely reflects the stronger correlations of mid-scale turbulence features with either the higher-frequency OAM phase patterns or the annular intensity envelope. These correlations may induce asymmetric phase distortions, which would reduce the value of inner products during sorting. 
	
	After a certain point, AT distortions become homogeneously high-frequency and speckle-like. These average out more easily during mode sorting integrations, and this would account for the XTP dips following the peaks. Finally, the rebounding increase in XTP at extreme values of $D/r_0$ is most likely due to the higher frequency AT features broadening the electric field at the detector. In other words, we obtain less phase information if AT-induced scintillation leads to greater amounts of high-frequency angular components being cut-off by our digital aperture.
	
	SNR results in Fig.~\ref{fig:5}(b) suggest similar trends to those shown for XTP, with cases having larger $\Delta\ell$, e.g.~(0, 18), also having the lowest SNR for most values of $D/r_0$, and vice-versa for cases with smaller $\Delta\ell$, e.g.~(0, 2). The highest SNR value of 100 is reached with the latter at $D/r_0 \sim 0.6$, but it decreases to 34 at $D/r_0 \sim 100$. With the (0, 18) case, at $D/r_0 \sim 7$ we reach the lowest SNR of 20, which started off as 71 for $D/r_0 \sim 0.01$. As for XTP, the largest SNR variance between different cases of ($\ell_1, \ell_2$) occurs for mid-level turbulence strengths where $D/r_0 \sim 6$.
	
	\begin{figure*}
		\centering
		\includegraphics[width=1\linewidth]{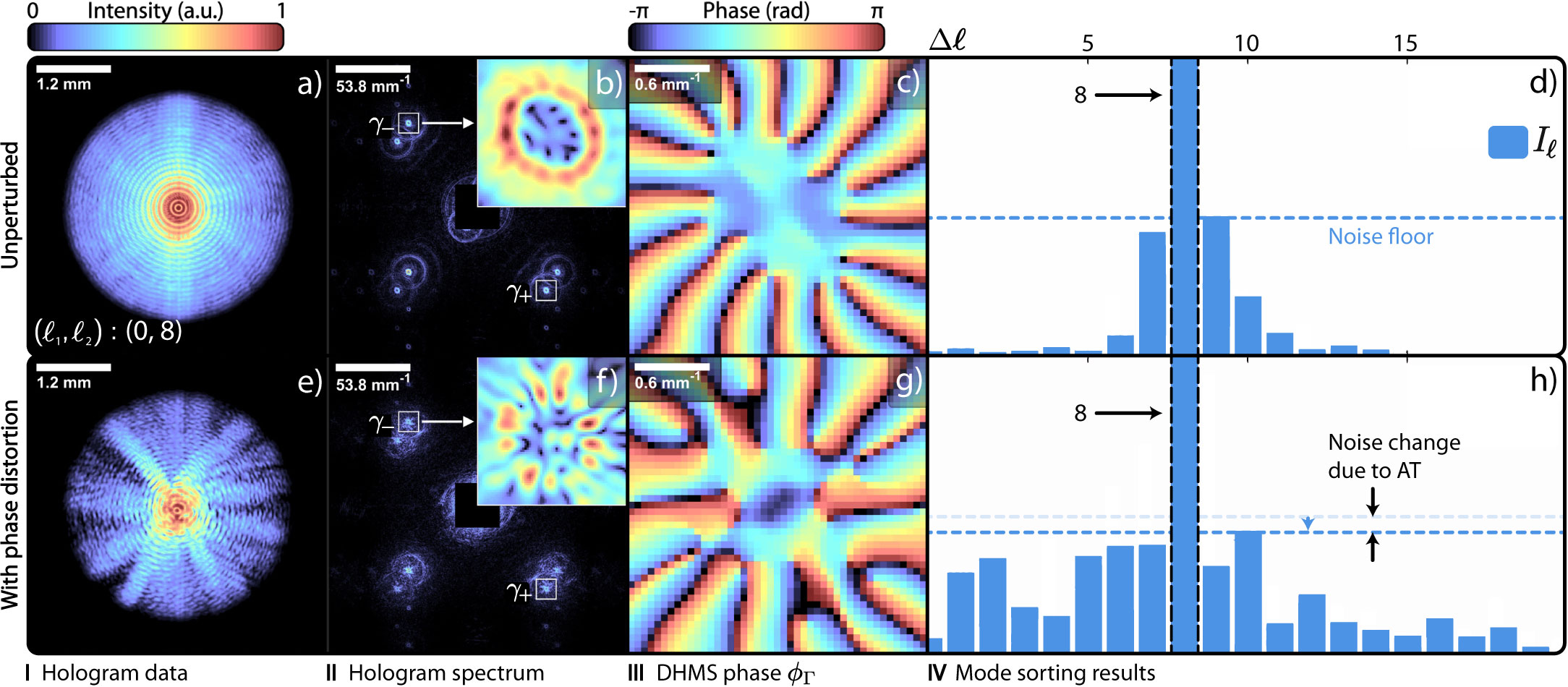}
		\caption[]{Experimental results of OAM DHMS for one mode pair: $(\ell_1,\ell_2) = (0,8)$. (a)-(d) and (e)-(h) show the results without and with the phase-distorting mask, respectively. Columns I, II, and III show the recorded hologram intensity profile, its spectrum, and the phase $\phi_{\Gamma}$ of the field $\Gamma$, respectively. Insets in column II show the intensity of $\gamma_-$. The bars in column IV show the DHMS amplitudes $I_{\ell}$ of the $\Gamma$ fields. Errors and noise floors for $I_{\ell}$ are also indicated using arrows and dashed lines.}
		\label{fig:7}
	\end{figure*}

\begin{figure*}
\centering
\includegraphics[width=1\linewidth]{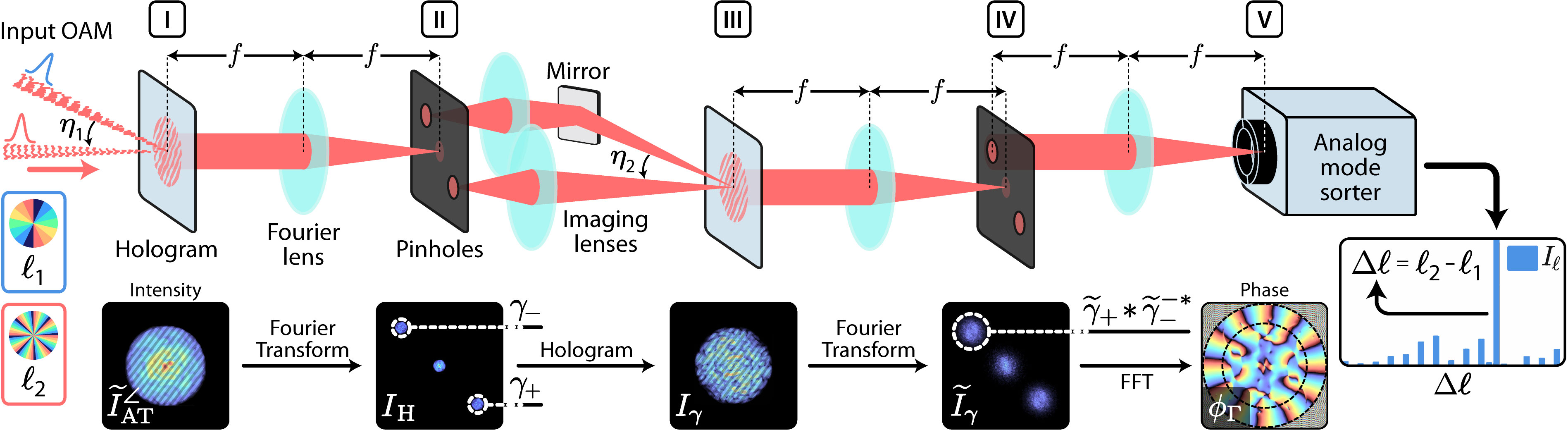}
\caption[]{Physical implementation of OAM-APSK analog holographic mode sorting (AHMS). Different steps in the algorithm and physical setup are denoted by the column numbers I through V. We assume that it is possible to easily record and read-out physical holograms (gray screens) using something like a diffusive screen or nonlinear material. I) The first hologram is recorded, as in regular DHMS, of the interference between two OAM modes. II) We take a Fourier transform of the hologram using a lens, and at the Fourier plane we sample the spectrum $I_h$ using two pinholes on the spots that correspond to $\gamma_-$ and $\gamma_+$. III) The latter two fields are imaged to record another hologram, $I_{\gamma}$. IV) We use a lens to project the spectrum $\widetilde{I}_{\gamma}$ onto a pinhole screen again, this time only filtering out the field $\widetilde{\gamma}_+ \ast \widetilde{\gamma}_-^{-*}$. V) A final lens takes the latter's Fourier transform, yielding $\gamma_+\gamma_-^*$, from which we can extract $\phi_{\Gamma}$ and use an analog mode sorter (e.g.~a wavefront sensor or cascaded spatial light modulators) to measure the OAM-APSK mode $\Delta\ell$.}
\label{fig:8}
\end{figure*}
	
	In summary, the XTP and SNR simulations demonstrate that both parameters remain at viable levels for a wide range of turbulence strengths. SNR never goes below 20 for APSK modes up to $\Delta\ell = 18$, and it only decreases by a factor of 5 even for very strong turbulence where $D/r_0 \sim 100$. Similarly XTP only increases by a factor of 5 and never exceeds 0.46 for a wide range of 21 modes being sampled. Overall, all our simulations suggest that DHMS provides a potentially 100\% signal recovery rate for APSK modes.
	

	\section{Experiment}
	
	To further verify the proposed approach, we carried out experiments with a setup schematically shown in Fig.~\ref{fig:6} (more details are given in Appendix Section III). Here, our goal is to provide a proof-of-concept for transferring a single OAM APSK bit. As such, we do not implement a full communication scheme as originally shown in Fig.~\ref{fig:1}. Instead, we substitute the pulsed laser with a continuous wave laser and refrain from implementing EOM gates for switching. For the phase-distorting mask, we used a 1-mm-thick polypropylene slab with a diameter of 3 cm., which approximates the effect of turbulent distortion with an equivalent AT strength $D/r_0$ ranging between 7 and 27 (more details are included in Appendix Section III). 
	
	
	Figure~\ref{fig:7} shows experimental OAM-APSK DHMS results for the case of ($\ell_1, \ell_2$) = (0, 8). In the absence of external perturbations to the beam profile, Fig.~\ref{fig:7}(a) shows the hologram data with a clean annular profile. Its spectrum in Fig.~\ref{fig:7}(b) shows the offset $\gamma_-$ and $\gamma_+$ fields, where the inset displays the details of $\gamma_-$ that also has an annular distribution. The slight distortion is introduced by the systematic errors in our setup that result primarily from certain physical imperfections on the beam splitter surfaces. In Fig.~\ref{fig:7}(c), the phase profile of $\phi_{\Gamma}$ exhibits the clear azimuthal phase pattern of $2\Delta\ell\theta$. Consequently, sorting results in Fig.~\ref{fig:7}(d) show a clear peak at $\Delta\ell = 8$. The side-lobes are due to the systematic errors as described above, which corresponds to a systematic XTP of 0.59 and a systematic SNR of 6.8.
	
	After introducing the phase-distorting element in our system, Fig.~\ref{fig:7}(e) now shows the hologram data with streaking across the annular profile. Similarly, Fig.~\ref{fig:7}(f) shows that the power spectrum is also distorted, with the original ring profile being spread amorphously into several intensity lobes. By strong contrast, Fig.~\ref{fig:7}(g) shows that the phase $\phi_{\Gamma}$ retains fairly faithfully the original unperturbed azimuthal profile. Consequently, DHMS produces a clear and strong peak for the APSK mode $\Delta\ell = 8$ with only a negligible change in the noise floor, as shown clearly in Fig.~\ref{fig:7}(h). As a result, the SNR decreases only by a small amount to 4.5, and the XTP increases to 0.79 accordingly. Similar DHMS results are obtained for different cases of ($\ell_1, \ell_2$), which are included in Appendix Section III.
	
	Overall, these experimental results show that OAM DHMS signals are robust to phase distortion, since the SNR fluctuates by only 11\% while still maintaining a clear peak at $\Delta\ell = 8$. Notably, we obtain these results even while using a phase screen that simulates intense AT distortions with $D/r_0$ ranging up to 27. Significant improvements can be made in the future, simply by resolving the systematic errors in our experimental setup.

	\section{Discussion}
	
	In summary, we have proposed OAM-APSK and DHMS as a combined encoding and decoding solution to address the challenge of atmospheric turbulence. We proved, both numerically and experimentally, that the proposed approach is able to significantly bypass AT-induced intensity noise by relying solely on phase information for encoding bits via OAM mode interference. Moreover, OAM-APSK can easily be extended to multi-valued bit encoding, and we can incorporate OAM-APSK together with existing spatial-mode division multiplexing schemes. We describe these multiplexing systems in Appendix Section IV.
	
	As for DHMS, its continued study may open up broader perspectives for spectral and computational turbulence mitigation techniques. We expect that DHMS could be incorporated with existing adaptive optics systems to further improve resolution, SNR, and bit-error rates. Indeed, adaptive optics systems can more easily reduce pointing errors and other low-frequency aberrations \cite{ren2014adaptive,ren2014adaptive2}, while APSK-DHMS can effectively bypass AT-induced high-frequency noise.
	
	Presently, the major limitation for APSK-DHMS lies in the speed of numerical processing and the computational limits imposed by requiring larger digital arrays for better resolution. However, in Figure \ref{fig:8} we introduce a possible analog holographic mode sorting (AHMS) framework for OAM-APSK. Making use of only lenses, holographic recording materials (or even diffuse projection screens), and pinhole filters, we can carry out the optical phase conjugation and multiplication operations that are integral to DHMS. In essence, we can transform the input field of OAM modes into the electric field $\gamma_+\gamma_-^*$ such that the analog mode sorter positioned at the end can operate on the phase $\phi_{\Gamma}$, retrieving the APSK mode $\Delta\ell$ just as is done via DHMS. Thus, with AHMS, we can take advantage of the faster analog conjugate mode sorting techniques that are carried out via SLMs and photodetectors \cite{gibson2004free}, while having already eliminated a significant amount of phase noise via the holographic optical conjugation.
	
	Future investigations will involve developing a rigorous DHMS theory, implementing a real-time long-distance OAM-APSK link, as well as investigating the viability of encoding quantum bits using the OAM charge difference. We hope that the proposed OAM-APSK and holographic mode sorting would open up a great avenue for OAM-based free-space optical communication that could ultimately resolve the challenge imposed by atmospheric turbulence.

	\subsection*{Acknowledgments}
	
	This research was done with support by grants from the National Science Foundation (NSF) (ECCS1810169, ECCS-1842691, EFMA-1641099).	\\
	
	\bibliography{sample}

	\clearpage
	\newpage
	
	\onecolumngrid
	\section*{Appendix}
	\renewcommand{\theequation}{A\arabic{equation}}
	\setcounter{equation}{0}
		
	\twocolumngrid
	\renewcommand{\thefigure}{A\arabic{figure}}
	\setcounter{figure}{0}
	\setcounter{section}{0}
	\section{Theory}
	
	\subsection{Digital holographic mode sorting derivation}
	
	Here we include a step-by-step derivation of the final theoretical result for the interference term $\overline{\gamma} = \gamma_+\gamma_-^*$, which contains the OAM azimuthal-phase shift keying (APSK) mode information obtained via digital holographic mode-sorting (DHMS).
	
	Using a spiral phase mask (with angular coordinate $\theta$), we define the $n^{\rm th}$ electric field $E_n$ (with amplitude $A_n$) of a beam propagating in the $z$ direction (with wave vector component $k_z$) with integer-valued OAM azimuthal mode $\ell_n$:
	\begin{equation}
	E_n = A_n e^{ik_z} e^{i\ell_n\theta}.
	\end{equation}
	Next, we assume that all fields start at $z = 0$ and have the same square-integrable amplitude function $A$. For any two modes $\ell_1$ and $\ell_2$, we begin the propagation right after the two coherent electric fields $E_1$ and $E_2$ are combined into a total field $E$:
	\begin{equation}
	\label{eq:3}
	E = E_1 + E_2 = A e^{ik_z0}(  e^{i\ell_1\theta} +  e^{i\ell_2\theta} ) = A (e^{i\ell_1\theta} +  e^{i\ell_2\theta}).
	\end{equation}
	
	Assuming that AT is only a phase perturbation $\Phi_{\mathrm{AT}}$ and can be represented by a very thin phase mask
	\begin{equation} 
	\Phi_{\mathrm{AT}}(x,y) = e^{i\phi_{\mathrm{AT}}(x,y)},
	\end{equation}
	we can describe the electric field $E_{\mathrm{AT}}$ after $\Phi_{\mathrm{AT}}$ as
	\begin{equation}
	\label{eq:5}
	\begin{split}
	E_{\mathrm{AT}} = E\cdot\Phi_{\mathrm{AT}} = A(\Psi_{\ell_1} + \Psi_{\ell_2})\Phi_{\mathrm{AT}},
	\end{split}
	\end{equation}
	where $\cdot$ denotes a pointwise product, and for shorthand we labeled the OAM phase functions with modes $\ell_1$ and $\ell_2$ as $\Psi_{\ell_1}$ and $\Psi_{\ell_2}$. 
	
	It is obvious that upon calculating the intensity of $E_{\mathrm{AT}}$ before propagation, the exponential phase $\phi_{\mathrm{AT}}$ is canceled out. However, for practical free-space propagation, $E_{\mathrm{AT}}$ will most often be propagated distances of several kilometers, which is equivalent to Fraunhofer propagation. Then at the receiver end, the field $\widetilde{E}_{\mathrm{AT}}$ is given by a coordinate-scaled Fourier (or Fraunhofer) transform $\mathcal{F}\{\cdot\}$ of $E_{\mathrm{AT}}$ \cite{goodman2005introduction}:
	\begin{equation}
	\begin{split}
	\mathcal{F}\{&E_{\mathrm{AT}}\} \\ =&\ \frac{e^{ikz}e^{i\frac{k}{2z}(\widetilde{x}^2 + \widetilde{y}^2)}}{i \lambda z} \int_{-\infty}^{\infty} E_{\rm AT} (x,y) e^{-i \frac{2\pi}{\lambda z} (x \widetilde{x} + y \widetilde {y})} \mathrm{d}x\mathrm{d}y,
	\end{split}
	\end{equation}
	where $\lambda$ is the wavelength, $z$ is the propagation distance, and we make the coordinate change ($x,y$) $\rightarrow$ ($\widetilde{x},\widetilde{y}$). For simplicity, we omit the effect of limiting apertures at both the source and receiver planes. Denoting the Fraunhofer transforms of $A$, $\Psi$, and $\Phi_{\mathrm{AT}}$ with over-script tildes, this gives
	\begin{equation}
	\begin{split}
	\label{eq:6}
	\widetilde{E}_{\mathrm{AT}} & = \mathcal{F}\{E_{\mathrm{AT}}\} = \mathcal{F}\{A(\Psi_{\ell_1} + \Psi_{\ell_2})\Phi_{\mathrm{AT}}\} \\
	& = \widetilde{\alpha}_1 \ast \widetilde{\Phi}_{\mathrm{AT}} + \widetilde{\alpha}_2 \ast \widetilde{\Phi}_{\mathrm{AT}},
	\end{split}
	\end{equation}
	where we refer to the convolutions $\widetilde{A} \ast \widetilde{\Psi}_{\ell_n}$ as OAM spectra $\widetilde{\alpha}_n$:
	\begin{equation}
	\widetilde{\alpha}_n = \widetilde{A} \ast \widetilde{\Psi}_{\ell_n}.
	\end{equation}
	Thus, Fraunhofer propagation is equivalent to convolving the noise spectrum separately with the two OAM spectra $\widetilde{\alpha}_1$ and $\widetilde{\alpha}_2$. 
	
	Analytically calculating the intensity of $\widetilde{E}_{\mathrm{AT}}$ is nontrivial, since there are convolution operations that must be done between fractional power functions and exponentials that contain radially varying arguments. However, instead of taking the intensity measured at the detector as our end-result, we use a digital holographic technique to extract useful phase information from $\widetilde{\alpha}_n$. We add a linear phase,
	\begin{equation}
	\phi(\eta) = k \sin(\eta)\tilde{x} = k_{\eta}\tilde{x},
	\end{equation}
	where $k$ is the wave vector magnitude and $\tilde{x}$ is the $x$-axis coordinate at the detector, by tilting one of the beams at an angle $\eta$ in the $x$-$z$ plane. Afterwards, we measure the electric field $\widetilde{E}_{\mathrm{AT}}^{\angle}$ and intensity $\widetilde{I}_{\mathrm{AT}}^{\angle}$ at the receiver:
	
	\begin{equation}
	\label{eq:7}
	\widetilde{E}_{\mathrm{AT}}^{\angle} = \widetilde{\alpha}_1 \ast \widetilde{\Phi}_{\mathrm{AT}} + \widetilde{\alpha}_2 \ast \widetilde{\Phi}_{\mathrm{AT}} \cdot e^{i\phi(\eta)},
	\end{equation}
	\begin{equation}
	\label{eq:8}
	\begin{split}
	\widetilde{I}_{\mathrm{AT}}^{\angle} = &\  |\widetilde{E}_{\mathrm{AT}}^{\angle}|^2 \\
	= &\ 
	|\widetilde{\alpha}_1 \ast \widetilde{\Phi}_{\mathrm{AT}}|^2 + |\widetilde{\alpha}_2 \ast \widetilde{\Phi}_{\mathrm{AT}}|^2 \\ &+ (\widetilde{\alpha}_1 \ast \widetilde{\Phi}_{\mathrm{AT}}) \cdot (\widetilde{\alpha}_2 \ast \widetilde{\Phi}_{\mathrm{AT}})^* \cdot e^{-i\phi(\eta)} \\ &+ 
	(\widetilde{\alpha}_1 \ast \widetilde{\Phi}_{\mathrm{AT}})^* \cdot (\widetilde{\alpha}_2 \ast \widetilde{\Phi}_{\mathrm{AT}}) \cdot e^{i\phi(\eta)}.
	\end{split}
	\end{equation}
	
	The principle of digital holography is based on using the fast-Fourier transform (FFT) to isolate high-frequency information, and its effect can be gauged by calculating $I_H$ or the \textit{inverse} Fraunhofer transform $\mathcal{F}^{-1}\{\cdot \}$ of the intensity $\widetilde{I}_{\mathrm{AT}}^{\angle}$:
	\begin{equation}
	\label{eq:9}
	\begin{split}
	I_{\rm H}  = &\ \mathcal{F}^{-1}\{\widetilde{I}_{\mathrm{AT}}^{\angle} \} \\
	= &\
	(\alpha_1 \cdot \Phi_{\mathrm{AT}}) \ast ( \alpha_1^{-} \cdot \Phi_{\mathrm{AT}}^{-} )^* \\ &+ 
	(\alpha_2 \cdot \Phi_{\mathrm{AT}}) \ast ( \alpha_2^{-} \cdot \Phi_{\mathrm{AT}}^{-} )^* \\ &+ 
	(\alpha_1 \cdot \Phi_{\mathrm{AT}}) \ast (\alpha_2^{-} \cdot \Phi_{\mathrm{AT}}^{-})^* \ast \delta(k_x-k_{\eta}) \\ &+ 
	(\alpha_1^- \cdot \Phi_{\mathrm{AT}}^-)^* \ast (\alpha_2 \cdot \Phi_{\mathrm{AT}}) \ast \delta(k_x+k_{\eta}),
	\end{split}
	\end{equation}
	where we make the coordinate change ($\widetilde{x},\widetilde{y}$) $\rightarrow$ ($k_x,k_y$), and the superscript minus signs indicate a sign inversion:
	\begin{equation}
	\alpha^-(x,y) = \alpha(-x,-y).
	\end{equation}
	For shorthand, we denote the sum of zero-frequency terms as $\gamma_0$ and the interference terms as $\gamma_+$ and $\gamma_-$:
	\begin{equation}
	\begin{split}
	\gamma_0 = &\ (\alpha_1 \cdot \Phi_{\mathrm{AT}}) \ast ( \alpha_1^{-} \cdot \Phi_{\mathrm{AT}}^{-} )^* \\ &+
	(\alpha_2 \cdot \Phi_{\mathrm{AT}}) \ast ( \alpha_2^{-} \cdot \Phi_{\mathrm{AT}}^{-} )^*, \\
	\gamma_+ = &\ (\alpha_1 \cdot \Phi_{\mathrm{AT}}) \ast (\alpha_2^{-} \cdot \Phi_{\mathrm{AT}}^{-})^*, \\
	\gamma_- = &\ (\alpha_1^- \cdot \Phi_{\mathrm{AT}}^-)^* \ast (\alpha_2 \cdot \Phi_{\mathrm{AT}}), \\
	I_{\rm H}  =  &\ \gamma_0 + \gamma_+ \ast \delta(k_x-k_{\eta}) + \gamma_- \ast \delta(k_x+k_{\eta}).
	\end{split}
	\end{equation}
	The result is that the linear phase imposed on one of the beams creates two interference terms in the frequency domain that are symmetrically placed on opposite sides of the origin along the $k_x$-axis. We are only interested in these two terms when filtering out the noise.
	
	Spectral noise is canceled when 1) either of the two $\gamma$ terms is point-wise multiplied by its sign-flipped version $\gamma^-$ or 2) when one $\gamma$ term is multiplied by the complex conjugate of the other. This gives the noise-filtered field $\overline{\gamma}$:
	\begin{equation}
	\begin{split}
	\overline{\gamma} = &\ \gamma_+\gamma_+^- = \gamma_+\gamma_-^* \\ = &\ [(\alpha_1 \cdot \Phi_{\mathrm{AT}}) \ast(\alpha_2^{-} \cdot \Phi_{\mathrm{AT}}^{-})^*] \\ & \cdot [(\alpha_1^{-} \cdot \Phi_{\mathrm{AT}}^{-}) \ast(\alpha_2 \cdot \Phi_{\mathrm{AT}})^*].
	\end{split}
	\label{eq:10}
	\end{equation}
	
	While the mathematics of this final expression do not give a final obvious result that the noise is canceled, one can intuit that after calculating $I_{\rm H}$, the AT terms revert to their non-spectral forms, i.e. from the Kolmogorov spectrum $\widetilde{\Phi}_{\mathrm{AT}}$ to the instantaneous phase $\Phi_{\mathrm{AT}}$ experienced by the beams during propagation. Unlike $\widetilde{\Phi}_{\mathrm{AT}}$ which contains phase \textit{and} amplitude information, $\Phi_{\mathrm{AT}}$ only contains \textit{phase} information, which cancels out when multiplied by its conjugate. This is exactly what happens in the final step to calculate $\overline{\gamma}$, and consequently we can use the latter's phase $\arg(\overline{\gamma}) = \phi_{\Gamma}$ to preserve OAM information through AT.
	
	By making a few simplifying assumptions to $\overline{\gamma}$, in the next two subsections we show how DHMS preserves twice the azimuthal phase difference while canceling out any turbulent noise. We only include either the OAM spectra or the AT phase noise in the DHMS propagations and calculations.
	
	\subsection{DHMS with no turbulence}
	
	In the main paper, we show that the final DHMS result, $\phi_{\Gamma}$, always contains the azimuthal phase $2\Delta\ell\theta$. This can be demonstrated by using a scenario where we are transmitting noiseless OAM modes and modifying the expression for $\overline{\gamma}$.
	
	We begin by setting the initial turbulent phase $\Phi_{\mathrm{AT}}$ to a matrix of ones, which greatly simplifies $\overline{\gamma}$ into $\overline{\gamma}_\ell$:
	\begin{equation}
	\overline{\gamma}_\ell = (\alpha_n \ast \alpha_m^{-*}) \cdot (\alpha_n^- \ast \alpha_m^*),
	\end{equation}
	where the subscripts $n$ and $m$ represent two separate OAM modes. Next, we can obtain very simple expressions in the derivation of $\overline{\gamma}_{\ell}$ by setting the amplitude $A$ of the OAM spectra to that of a Bessel (or non-diffracting) beam:
	\begin{equation}
	\alpha_n = J_{\ell_n}(k_r r)e^{i\ell_n\theta} = f(r)f(\theta),
	\end{equation}
	where $J_{\ell_n}$ is an $\ell_n^{\rm th}$ order Bessel function of the first kind, $k_r$ is the radial wave vector component, and $f(r)$ and $f(\theta)$ are separable radial and azimuthal functions (respectively) of $\alpha_n$.
	
	We compute the convolutions by using the convolution property of the Fourier transform:
	\begin{equation}
	\alpha_n \ast \alpha_m^{-*} = \mathcal{F}^{-1}\{\mathcal{F}\{\alpha_n\} \cdot \mathcal{F}\{\alpha_m^{-*}\} \} =  \mathcal{F}^{-1}\{\widetilde{\alpha}_n \cdot \widetilde{\alpha}_m^* \}.
	\end{equation}
	Thanks to the separability of $\alpha_n$ in polar coordinates, the Fourier transform can be expressed as an infinite sum of weighted Hankel transforms $\mathcal{H}_p\{\cdot\}$ of order $p$ \cite{goodman2005introduction},
	\begin{equation}
	\mathcal{H}_p\{f(r)\} = 2\pi\int^{\infty}_{0} r f(r) J_p(2\pi r \rho)\mathrm{d}r,
	\end{equation} 
	and this can be used to calculate $\widetilde{\alpha}_n$ from $\alpha_n$ \cite{goodman2005introduction}:
	\begin{equation}
	\widetilde{\alpha}_n = \sum^{\infty}_{p = -\infty} c_p(-i)^p e^{ip\zeta} \mathcal{H}_p\{f(r)\},
	\end{equation}
	where $\zeta$ is the frequency-space angular coordinate, and $c_p$ is an azimuthal Fourier series coefficient \cite{goodman2005introduction}:
	\begin{equation}
	c_p = \frac{1}{2\pi}\int^{2\pi}_{0}f(\theta)e^{-i p \theta}\mathrm{d}\theta.
	\end{equation}
	
	Starting with $c_p$, we substitute $f(\theta)$ into the integral:
	\begin{equation}
	c_p = \frac{1}{2\pi}\int^{2\pi}_{0}e^{i\ell_n\theta}e^{-i p \theta}\mathrm{d}\theta = \frac{1}{2\pi}\int^{2\pi}_{0}e^{i(\ell_n - p) \theta}\mathrm{d}\theta.
	\end{equation}
	Straight away, it is evident that $c_p$ is zero unless we set $p = \ell_n$. This reduces $c_p$ to a constant value of one,
	\begin{equation}
	c_{p}\rvert_{p = \ell_n} = \frac{1}{2\pi}\int^{2\pi}_0 e^{i(\ell_n - \ell_n)\theta} \mathrm{d}\theta = \frac{1}{2\pi}\int^{2\pi}_0 \mathrm{d}\theta = 1,
	\end{equation}
	and greatly simplifies the expression for $\alpha_n$:
	\begin{equation}
	\widetilde{\alpha}_n = (-i)^{\ell_n} e^{i\ell_n\zeta} \mathcal{H}_{\ell_n}\{f(r)\}.
	\end{equation}
	
	Next, we can expand the Hankel transform:
	\begin{equation}
	\mathcal{H}_{\ell_n}\{f(r)\} = 2\pi\int^{\infty}_0 J_{\ell_n}(k_r r) J_{\ell_n}(2 \pi r \rho) r \mathrm{d}r,
	\end{equation}
	where we can simplify this expression using the orthogonality property of Bessel functions \cite{Vaity:15}:
	\begin{equation}
	\int^{\infty}_0 J_{\nu}(a t) J_{\nu}(b t) t \mathrm{d}t = \frac{\delta(a - b)}{a}.
	\end{equation}
	Using this identity, the Hankel transform reduces to a scaled annular delta function:
	\begin{equation}
	\mathcal{H}_{\ell_n}\{f(r)\} = 2\pi \frac{\delta(k_r - 2\pi\rho)}{k_r} = \frac{ \delta(\rho - k_r/ 2\pi)}{k_r}.
	\end{equation}
	Now we have the final expression for $\widetilde{\alpha}_n$:
	\begin{equation}
	\widetilde{\alpha}_n = (-i)^{\ell_n} e^{i\ell_n\zeta} \frac{ \delta(\rho - \rho_r)}{k_r},
	\end{equation}
	where $\rho_r = k_r/ 2\pi$.
	
	The main takeaway from the last expression is that OAM is conserved after Fourier transformation, which also results in another annular intensity distribution. Moreover, the azimuthally varying exponential phase $f(\theta)$ functions as a kind of spatial filter, where the azimuthal charge $\ell_n$ determines the Hankel transform order $p$ and consequently the amplitude $f(r)$ of the angular spectrum. As an aside, using a Bessel beam as the amplitude basis for OAM leads to a far-field intensity distribution whose size is independent of the azimuthal charge $\ell_n$. In our simulations and experiments, we primarily used an initial top-hat or circ function as the beam amplitude, which is not optimal since the far-field intensity diameter then depends on the azimuthal mode. This variance in far-field spectral diameter leads to problems in DHMS, where results are increasingly worse with larger values of $\Delta\ell$ due to poor mode overlap. However, these problems could be solved by using a Bessel beam amplitude basis instead.
	
	We can now return to our initial goal of calculating the convolution of two independent OAM modes, substituting the new expressions for $\widetilde{\alpha}_n$:
	\begin{equation}
	\begin{split}
	\alpha_n \ast & \alpha_m^{-*} \\ = &\ \mathcal{F}^{-1}\{\widetilde{\alpha}_n \cdot \widetilde{\alpha}_m^*\} \\
	= &\ \mathcal{F}^{-1} \left\{ (-i)^{\ell_n} e^{i\ell_n\zeta} \frac{ \delta(\rho - \rho_r)}{k_r} \cdot i^{\ell_m} e^{-i\ell_m\zeta} \frac{ \delta(\rho - \rho_r)}{k_r} \right \} \\
	= &\ \mathcal{F}^{-1} \left\{ (-i)^{\ell_n} i^{\ell_m} e^{i(\ell_n- \ell_m)\zeta} \frac{ \delta(\rho - \rho_r)}{k_r} \right \} \\
	= &\ (-i)^{\ell_n} i^{\ell_m} \mathcal{F}^{-1} \left\{  e^{i\Delta\ell\zeta} \frac{ \delta(\rho - \rho_r)}{k_r} \right \},
	\end{split}
	\end{equation}
	where to simplify further analysis we approximate $\delta^2(\cdot)/k_r^2$ as $\delta(\cdot)/k_r$. It is evident that the APSK mode $\Delta\ell = \ell_n - \ell_m$ is obtained from this initial product of OAM spectra. From here, we can go through the same process that we used to derive $\widetilde{\alpha}_n$ in order to calculate the inverse Fourier transform, for which we can guess that the azimuthal mode $\Delta\ell$ is preserved again.
	
	Again we begin by calculating $c_p$,
	\begin{equation}
	c_p = \frac{1}{2\pi}\int^{2\pi}_{0}e^{i\Delta\ell\zeta}e^{-i p \theta}\mathrm{d}\zeta,
	\end{equation}
	which is reduced to one only in the case that $p = \Delta\ell$. We then calculate the Hankel transform of order $\Delta\ell$:
	\begin{equation}
	\begin{split}
	\mathcal{H}_{\Delta\ell}\{f(r)\} = &\ 2\pi\int^{\infty}_0 \frac{\delta(\rho - \rho_r)}{k_r} J_{\Delta\ell}(2 \pi r \rho) \rho \mathrm{d}\rho \\
	= &\ \frac{2\pi}{k_r} J_{\Delta\ell}(2\pi r \rho_r) \rho_r \\
	= &\ J_{\Delta\ell}(k_r r).
	\end{split}
	\end{equation}
	The final expression for the convolution $\alpha_n \ast \alpha_m^{-*}$ is then given by combining $c_p$ and $\mathcal{H}_{\Delta\ell}\{f(r)\}$:
	\begin{equation}
	\alpha_n \ast \alpha_m^{-*} = (-i)^{\ell_n}i^{\ell_m}(-i)^{\Delta\ell} e^{i\Delta\ell\theta} J_{\Delta\ell}(k_r r).
	\end{equation}
	
	In order to calculate $\overline{\gamma}_{\ell}$ using the last expression, we have to carry out a \textit{sign inversion} (or 180 degree rotation) on it. Owing to the radial symmetry in the Bessel function $J_{\Delta\ell}(k_rr)$ of $\alpha_n \ast \alpha_m^{-*}$, the inversion returns an identical function. Although the azimuthal phase term $\exp(i\Delta\ell\theta)$ does not have radial symmetry, a rotation at any arbitrary angle simply applies a global phase shift on it and would not affect the function at all. This can also be seen by inverting the arguments of the arctan2 function:
	\begin{equation}
	\theta^- = \arctan2(-y/-x) = \arctan2(y/x) = \theta.
	\end{equation}
	Therefore, both convolution terms in $\overline{\gamma}_{\ell}$ are identical:
	\begin{equation}
	\alpha_n \ast \alpha_m^{-*} = \alpha_n^- \ast \alpha_m^*.
	\end{equation}
	With this in mind, we finally have an expression for the undistorted OAM-APSK signal $\overline{\gamma}_{\ell}$:
	\begin{equation}
	\begin{split}
	\overline{\gamma}_{\ell} = &\ (\alpha_1 \ast \alpha_2^{-*})^2 \\
	= &\ (-i)^{2\ell_n}i^{2\ell_m}(-i)^{2\Delta\ell} e^{2i\Delta\ell\theta} J_{\Delta\ell}^2(k_r r) \\
	= &\ \kappa_i e^{2i\Delta\ell\theta} J_{\Delta\ell}^2(k_r r),
	\end{split}
	\end{equation}
	where we combine values of $i$ as one constant $\kappa_i$, and we clearly see the azimuthal phase $2\Delta\ell\theta$.
	
	\subsection{DHMS with only turbulence}
	
	In order to show how digital phase conjugation in the DHMS algorithm removes turbulent phase noise, we now describe a case where we transmit plane waves, instead of OAM beams, through a turbulent atmosphere. By including only the AT noise, this simplifies $\overline{\gamma}$ into $\overline{\gamma}_{\rm AT}$:
	\begin{equation}
	\overline{\gamma}_{\rm AT} = (\Phi_{\mathrm{AT}} \ast \Phi_{\mathrm{AT}}^{-*}) \cdot (\Phi_{\mathrm{AT}}^- \ast \Phi_{\mathrm{AT}}^*).
	\end{equation}
	Right away we can see that when the two functions being convolved are identical---unlike in the case with no noise and two different OAM modes---sign inversion and complex conjugation are \textit{equivalent}. In order to emphasize this point, we simply expand the convolutions.
	
	We start with the first convolution, using the original definition for $\Phi_{\mathrm{AT}}$:
	\begin{equation}
	\begin{split}
	\Phi_{\mathrm{AT}} \ast \Phi_{\mathrm{AT}}^{-*} = &\ e^{i\phi_{\mathrm{AT}}(x,y)} \ast e^{-i\phi_{\mathrm{AT}}(-x,-y)} \\
	= &\ \int e^{i\phi_{\mathrm{AT}}(x-x',y-y')} e^{-i\phi_{\mathrm{AT}}(-x',-y')} \mathrm{d} x' \mathrm{d} y'.
	\end{split}
	\end{equation}
	The next convolution is very similar:
	\begin{equation}
	\begin{split}
	\Phi_{\mathrm{AT}}^- \ast \Phi_{\mathrm{AT}}^{*} = &\ e^{-i\phi_{\mathrm{AT}}(x,y)} \ast e^{i\phi_{\mathrm{AT}}(-x,-y)} \\
	= &\ \int e^{-i\phi_{\mathrm{AT}}(x-x',y-y')} e^{i\phi_{\mathrm{AT}}(-x',-y')} \mathrm{d} x' \mathrm{d} y' \\
	= &\ (\Phi_{\mathrm{AT}} \ast \Phi_{\mathrm{AT}}^{-*})^*,
	\end{split}
	\end{equation}
	consequently:
	\begin{equation}
	\overline{\gamma}_{\rm AT} = ( \Phi_{\mathrm{AT}}^- \ast \Phi_{\mathrm{AT}}^{*} )^* \cdot ( \Phi_{\mathrm{AT}}^- \ast \Phi_{\mathrm{AT}}^{*} ) = \left| \Phi_{\mathrm{AT}}^- \ast \Phi_{\mathrm{AT}}^{*} \right|^2.
	\end{equation}
	
	No matter what the final expression of $\Phi_{\mathrm{AT}}^- \ast \Phi_{\mathrm{AT}}^{*}$ looks like, we can always reduce it to separable functions for amplitude and phase. Therefore, in any case, the exponential phase due to AT will always be eliminated in the final DHMS result. We reiterate that although we assumed plane waves instead of OAM modes as our signals, simulation and experimental results shown in the main paper---where most noise due to AT is mitigated---support the theoretical conclusion we lay out here.

	%
	
	\section{Simulation}
	
	\subsection{Atmospheric turbulence model}
	
	\begin{figure}
		[!hb]
		\centering
		\includegraphics[width=1\linewidth]{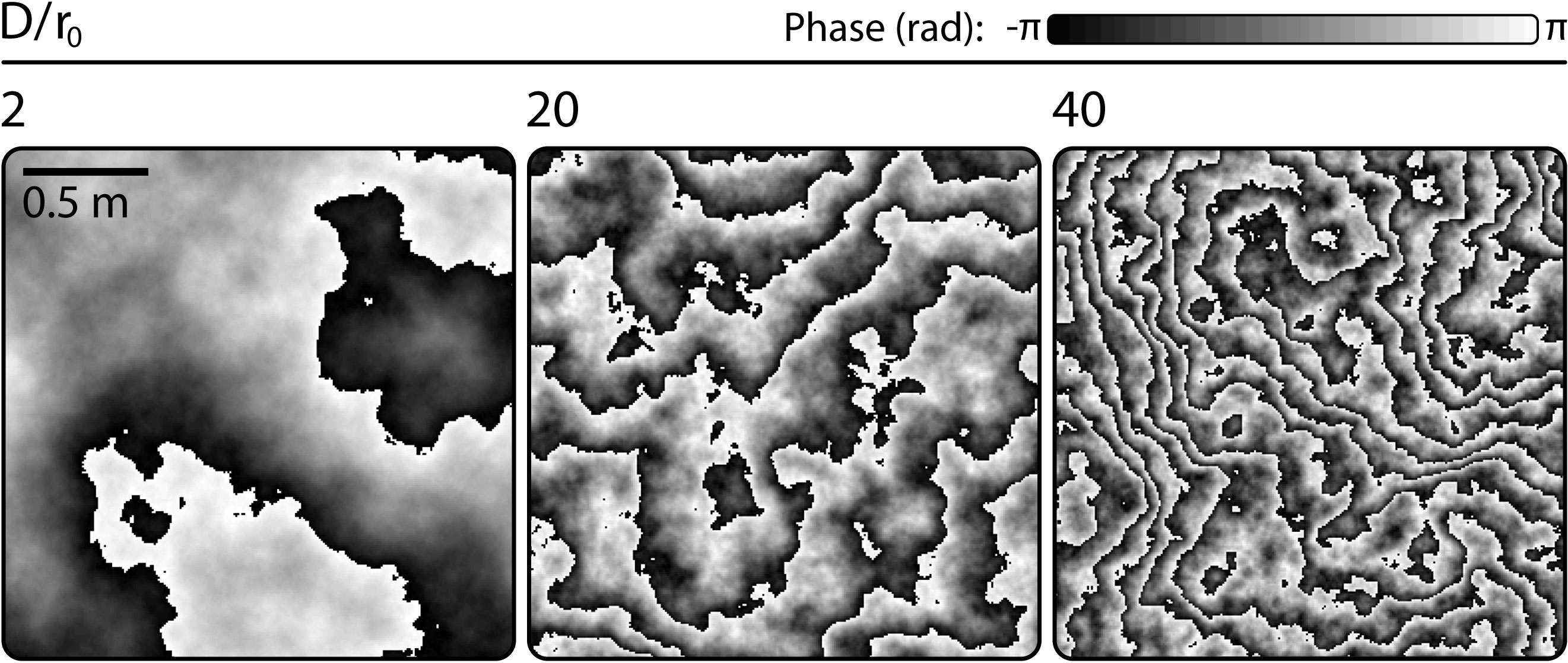}
		\caption[]{Examples of phase screens simulated using our model of atmospheric turbulence, with $D/r_0$ values of 2, 20, and 40 from left to right. In these simulations, the window diameter $D$ was 2 m., the number of grid points were $256^2$, the outer scale constant was 100 m, and the inner scale constant was 1 cm.}
		\label{fig:atex}
	\end{figure}
	\begin{figure}
		[!hb]
		\centering
		\includegraphics[width=1\linewidth]{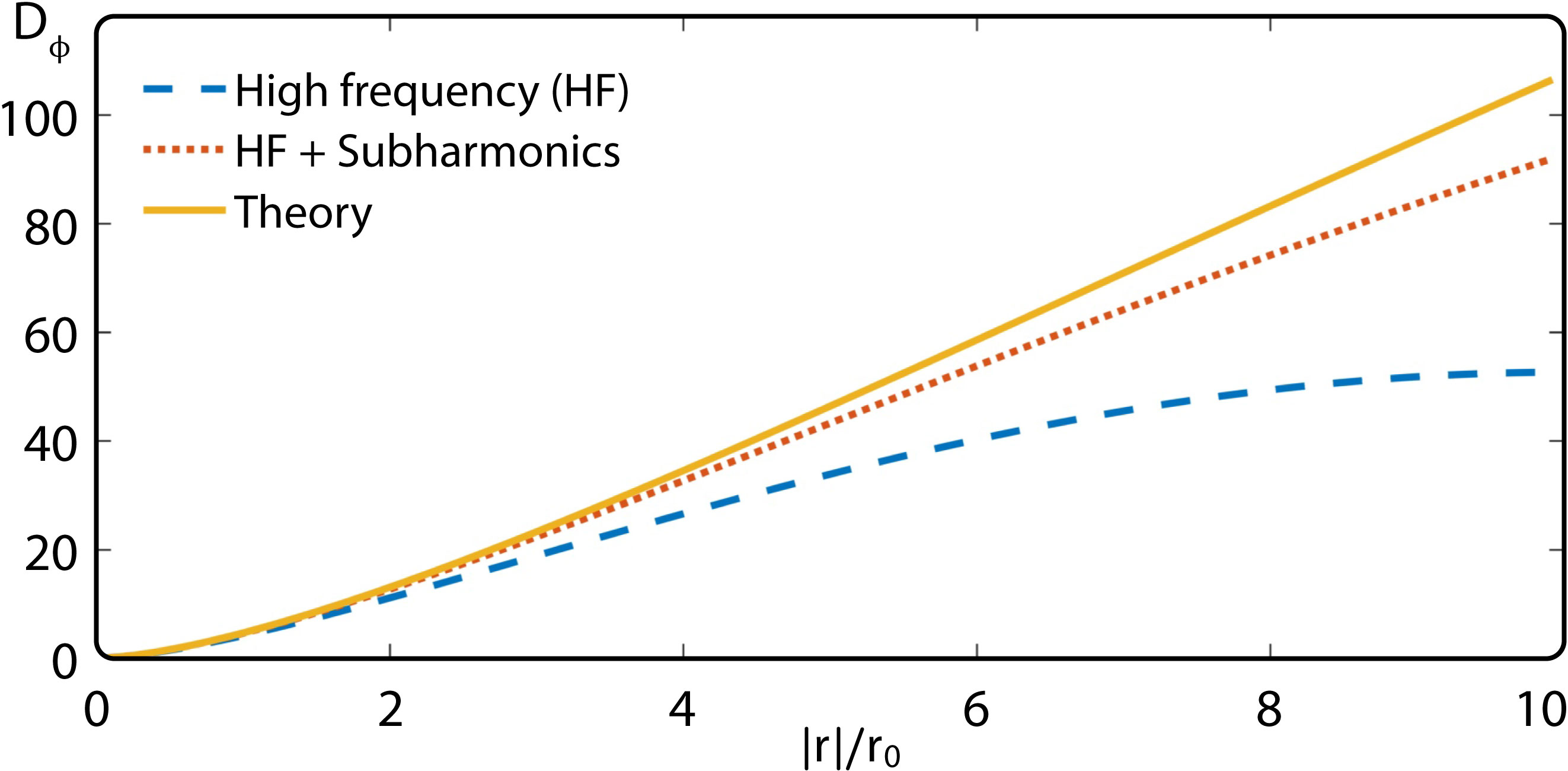}
		\caption[]{Plots of mean structure function constants $D_{\phi}$ for AT phase versus the separation in space $|r|/r_0$, where $r$ is normalized by the Fried coherence diameter. As expected, the purely high-frequency AT phase (blue) less closely approximates the theory (yellow), but the AT phase that includes subharmonics (red) approximates the theory very well. These results can also be compared to those found in Ref. \cite{johansson1994simulation}.}
		\label{fig:atstruct}
	\end{figure}
	We create a continuous atmospheric turbulence (AT) phase screen $\Phi_{\mathrm{AT}} = \exp(i \phi_{\mathrm{AT}})$ by calculating an FFT of the product between the turbulent 2D power spectrum $\widetilde{\phi}_{\mathrm{AT}}$ (which is the modified von-Karman spectrum in all simulations) and an array $c_{n,m}$ of zero-mean unit-variance complex-valued Gaussian white noise \cite{johansson1994simulation}:
	\begin{equation}
	\phi_{\mathrm{AT}}(\mathbf{r}) = \mathrm{FFT}\left\{ c_{n,m}(\mathbf{k}) \sqrt{\widetilde{\phi}_{\mathrm{AT}}(\mathbf{k})}  \right\},
	\end{equation}
	where $\mathbf{r}$ and $\mathbf{k}$ are vectors containing the space and spatial frequency coordinates ($x,y$) and ($k_x,k_y$), respectively.
	
	We verified the integrity of our AT model by numerically calculating the phase structure function $D_{\phi}$ and comparing it to the theoretical $D_{\phi}^{\mathrm{vK}}$ for a windowed AT phase mask made using a von-Karman spectrum. The latter function is given by \cite{schmidt2010numerical}:
	\begin{equation}
	D_{\phi}^{\mathrm{vK}} = 6.16 r_0^{-5/3} \left[ \frac{3}{5} k_0^{-5/3}  - \frac{(r/2 k_0)^{5/6}}{\Gamma(11/6)} K_{5/6} (k_0 r) \right],
	\label{eq:atsfxn}
	\end{equation}
	where $r_0$ is the Fried coherence diameter, $r$ is a separation distance, $k_0$ is the frequency for the outer scale constant $L_0$, $\Gamma$ is the gamma function $\Gamma(n) = (n-1)!$, and $K_{\nu}$ is the modified Bessel function of the second kind with order $\nu$. Numerically we used the identity relating $D_{\phi}$ to the autocorrelation $\Lambda$:
	\begin{equation}
	D_{\phi}^{\mathrm{vK}}(r) = 2\left[ \Lambda(0) - \Lambda(r) \right].
	\end{equation}
	We include examples of AT phase screens that we generated in Fig.~\ref{fig:atex}, and we show plots of $D_{\phi}(r)$ versus $r$ in Fig.~\ref{fig:atstruct}.
	
	Additionally, we included a subharmonic routine in order to account for tilt and other lower-order aberrations induced by AT \cite{lane1992simulation}. The final phase screen $\Phi_{\mathrm{AT}}$ was taken as the complex exponential of either the real or imaginary part of the array $\phi_{\mathrm{AT}}$ generated by the FT routine. 
	
	\subsection{Digital holography sampling}
	
	\begin{figure*}
		\centering
		\includegraphics[width=1\linewidth]{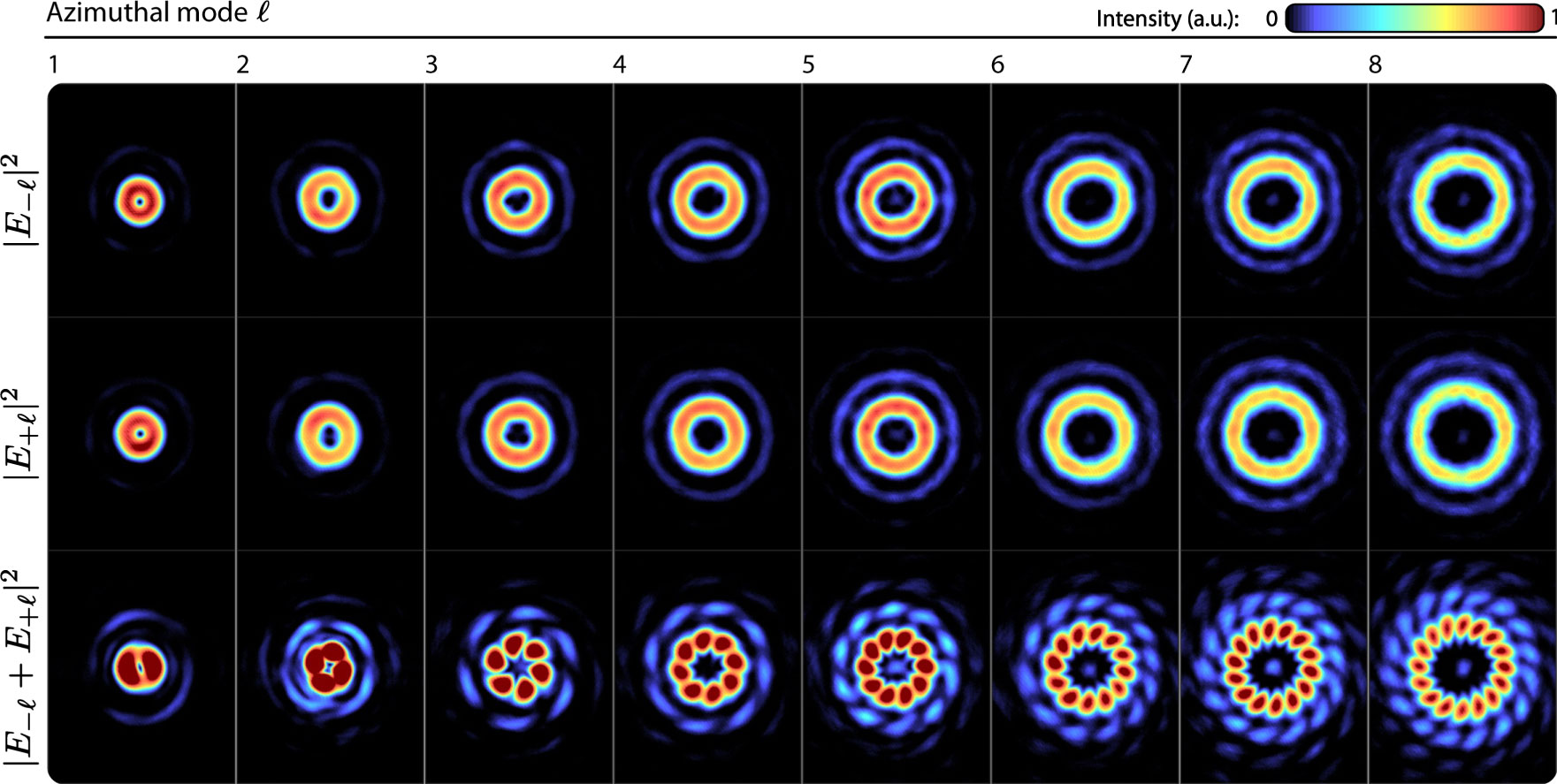}
		\caption[]{Initial intensity profiles of unperturbed OAM beams used in experiments. Columns separate images by OAM azimuthal modes $\ell$ from 1 through 8. The top, middle, and bottom rows show the intensity profiles of negative parity $(-\ell)$ azimuthal modes, positive parity $(+\ell)$ modes, and interference patterns between opposite parity modes, respectively.}
		\label{fig:oamex}
	\end{figure*}
	
	During the digital holography step of the DHMS simulation, only one field is tilted by an angle $\eta$ (or equivalently multiplied by a linear phase array). Crucially, $\eta$ must not exceed a maximum value $\eta_{\mathrm{max}}$, which is defined by the detector pixel size $\Delta x$ or Nyquist frequency \cite{cuche2000spatial}:
	\begin{equation}
	\eta \leq \eta_{\mathrm{max}} = \arcsin\left(\frac{\lambda}{2\Delta x}\right).
	\end{equation}
	
	When sampling the spectrum for $\phi_{\Gamma}$, we find that the useful portion of phase, after being offset due to the hologram, is constrained by the propagation parameters. Specifically, we only sample an area limited by the radius $k_{\rm max}$ in frequency space:
	\begin{equation}
	k_{\rm max} = \frac{D}{z_i \lambda},
	\end{equation}
	where $D$ is the initial beam diameter and $z_i$ is the propagation distance. Therefore this radius must also be taken into account when deciding what value to use for $\eta$.
	
	\subsection{Regression model for simulation results}
	
	Nonlinear polynomial regression was used for fitting curves to the crosstalk percentage (XTP), signal-to-noise ratio (SNR), and mode-specific crosstalk $\left<s_{\Delta}\right>$ simulation results. All data displayed logarithmic behavior in the x-dimension, so we used the logarithm of turbulence strength $\log_{10}(D/r_0)$ as the coordinates during fitting. If data also displayed logarithmic behavior in the y-dimension, then we used $\log_{10}(y)$ for the fits, and reverted to the original linear base for plotting.
	
	\begin{figure}
		[!ht]
		\centering
		\includegraphics[width=1\linewidth]{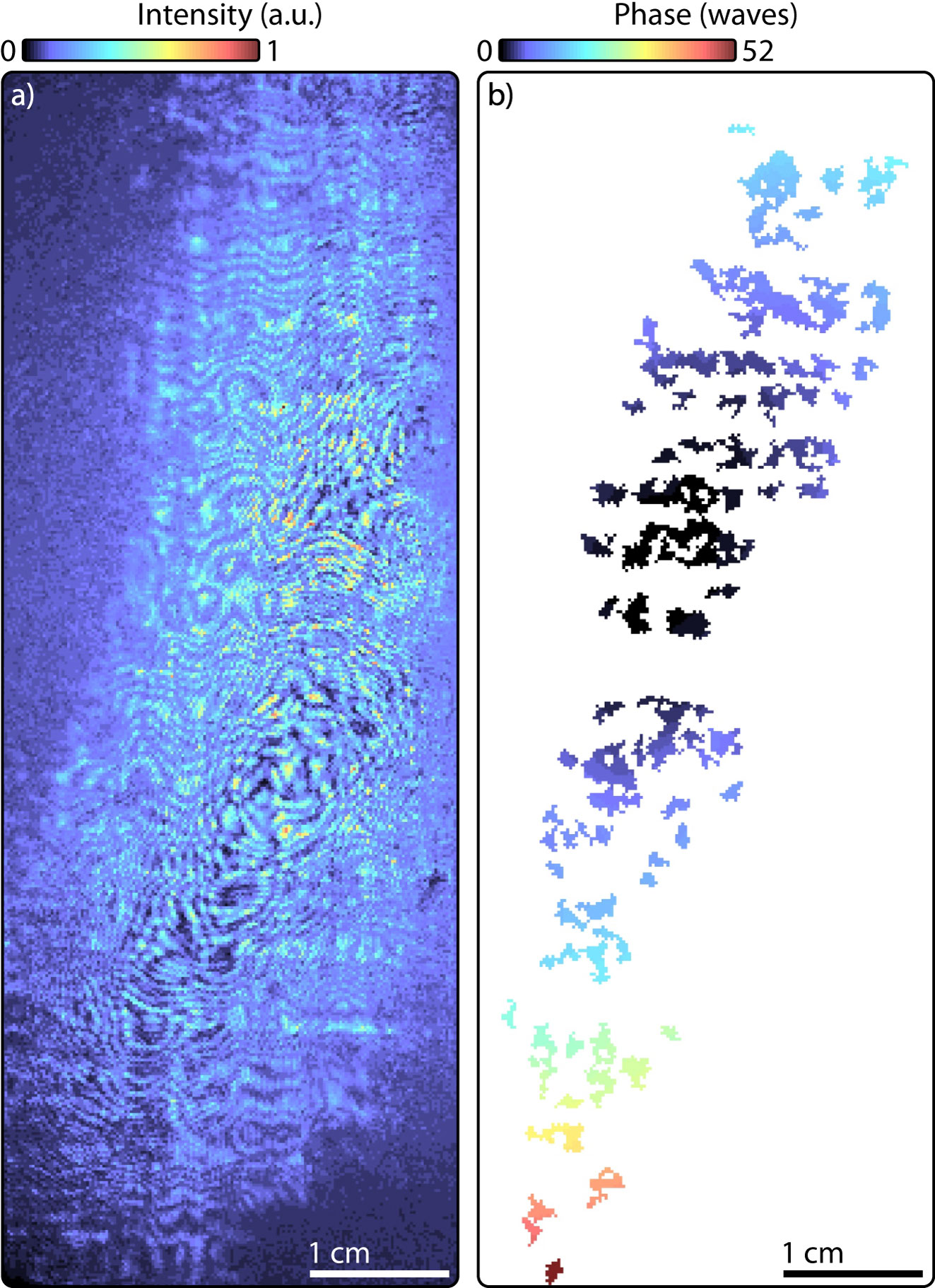}
		\caption[]{Intensity (a) and phase (b) measurements of experimental phase-distorting screen obtained using a phase-shifting Fizeau interferometer. The large variation in the surface is evident from the distorted interferogram and the significant amounts of fall-off error in the phase measurement, which ranges to at least 52 waves (at 633 nm).}
		\label{fig:verifire}
	\end{figure}
	
	\begin{figure}
		[!ht]
		\centering
		\includegraphics[width=1\linewidth]{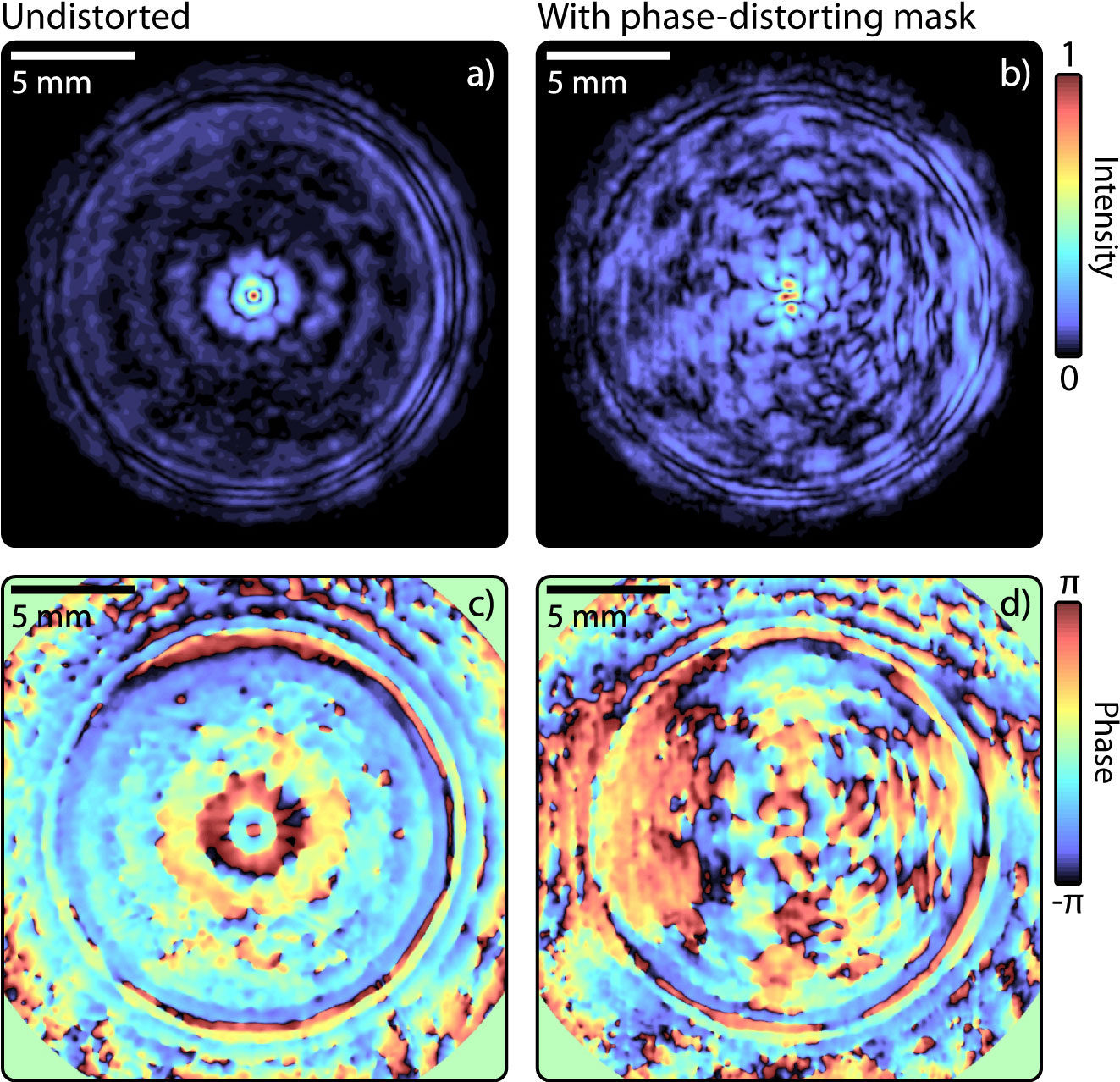}
		\caption[]{Digital holographic intensity and phase measurements of clean input wavefront before (a, c) and after (b, d) inserting a phase-distorting mask.}
		\label{fig:dhmsdr0}
	\end{figure}
	
	\begin{figure}
		[!ht]
		\centering
		\includegraphics[width=1\linewidth]{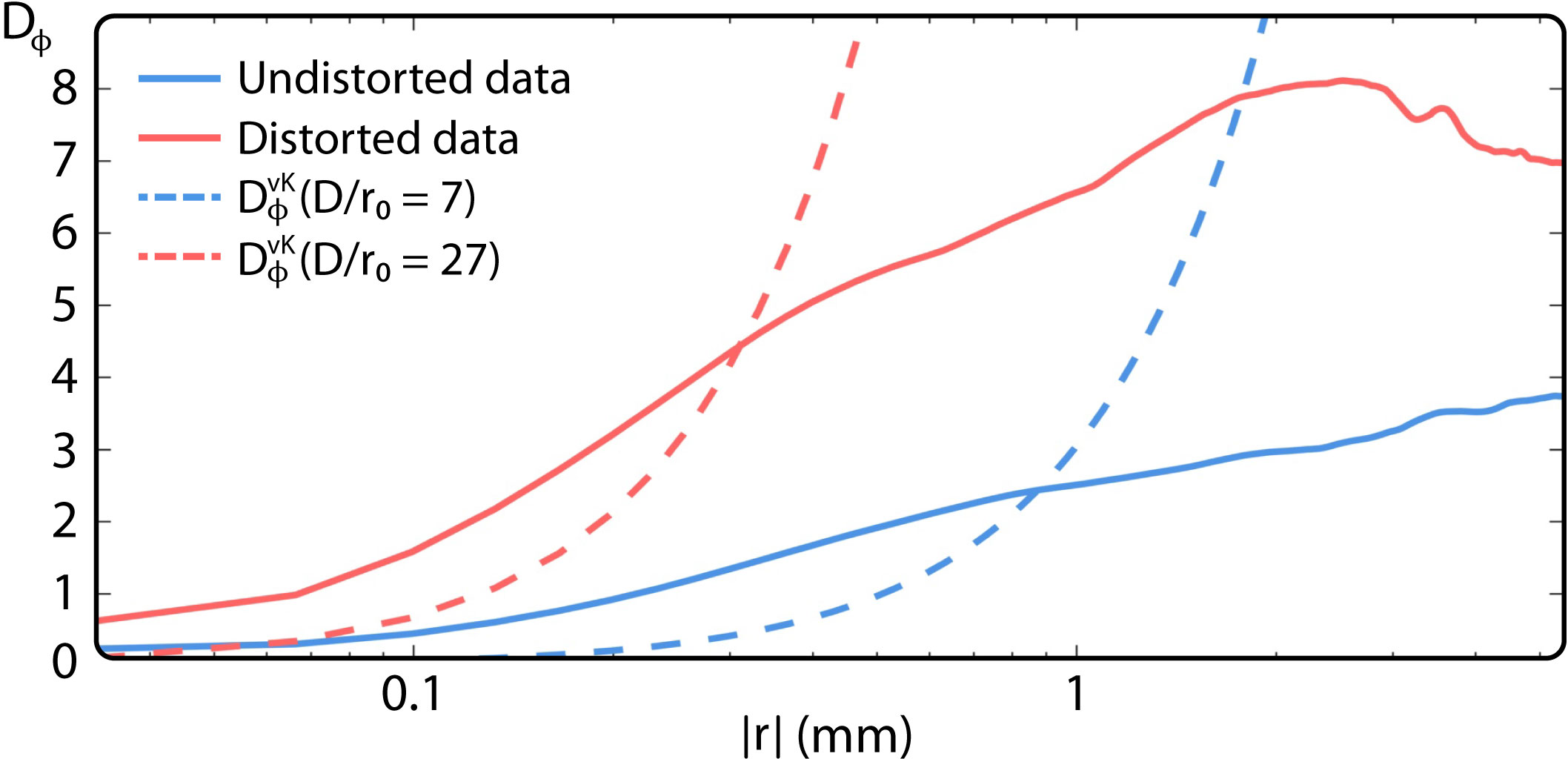}
		\caption[]{Phase structure function $D_{\phi}$ calculations from undistorted and distorted phase surfaces shown in Fig.~\ref{fig:dhmsdr0}(c, d), respectively. Red and blue curves correspond to the distorted and undistorted cases, respectively. Dashed lines correspond to theoretical estimates (Eq.~\ref{eq:atsfxn}) of the von-Karman phase structure functions $D_{\phi}^{vK}$, which were fit to the data-derived $D_{\phi}$ by ensuring that $D_{\phi}^{vK}$ are approximate upper bounds of $D_{\phi}$.}
		\label{fig:dhmsdr0-sfxn}
	\end{figure}
	
	\begin{figure*}
		[!ht]
		\centering
		\includegraphics[width=1\linewidth]{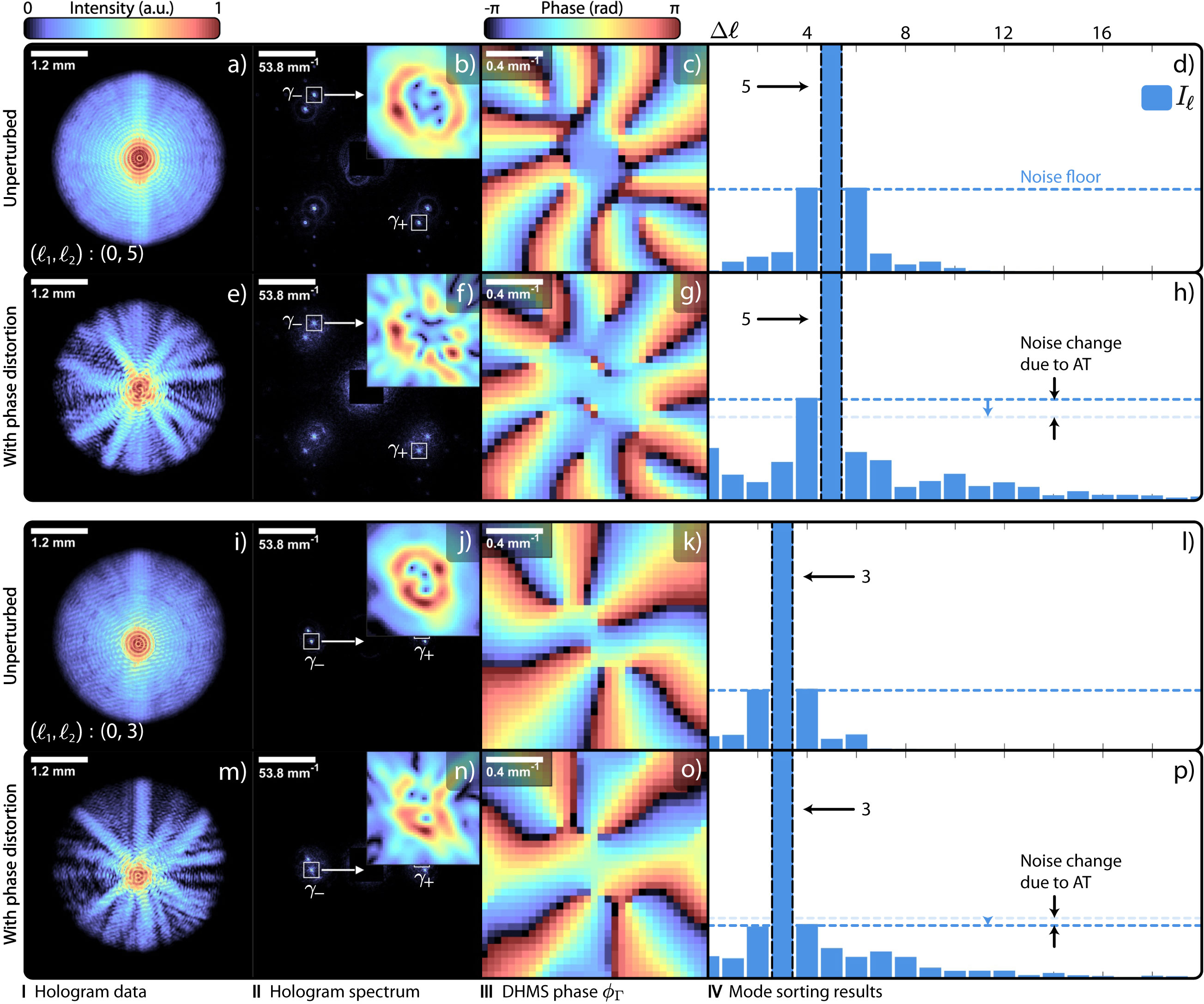}
		\caption[]{Additional experimental results of OAM-APSK DHMS. Two sets of OAM mode pairs, $(\ell_1,\ell_2) = \{(0,5),(0,3)\}$, are shown in parts (a)-(h) and (i)-(p), respectively. Columns I, II, III, and IV, show the hologram intensity, the subsequent spectrum, the phase $\phi_{\Gamma}$, and the DHMS results, respectively. Middle column insets show the intensity and phase of the uncorrected field $\gamma_-$. Unperturbed and phase-distorted results are grouped in sets of two rows for each case of ($\ell_1, \ell_2$).}
		\label{fig:oamres}
	\end{figure*}
	
	\begin{figure}
		[!ht]
		\centering
		\includegraphics[width=1\linewidth]{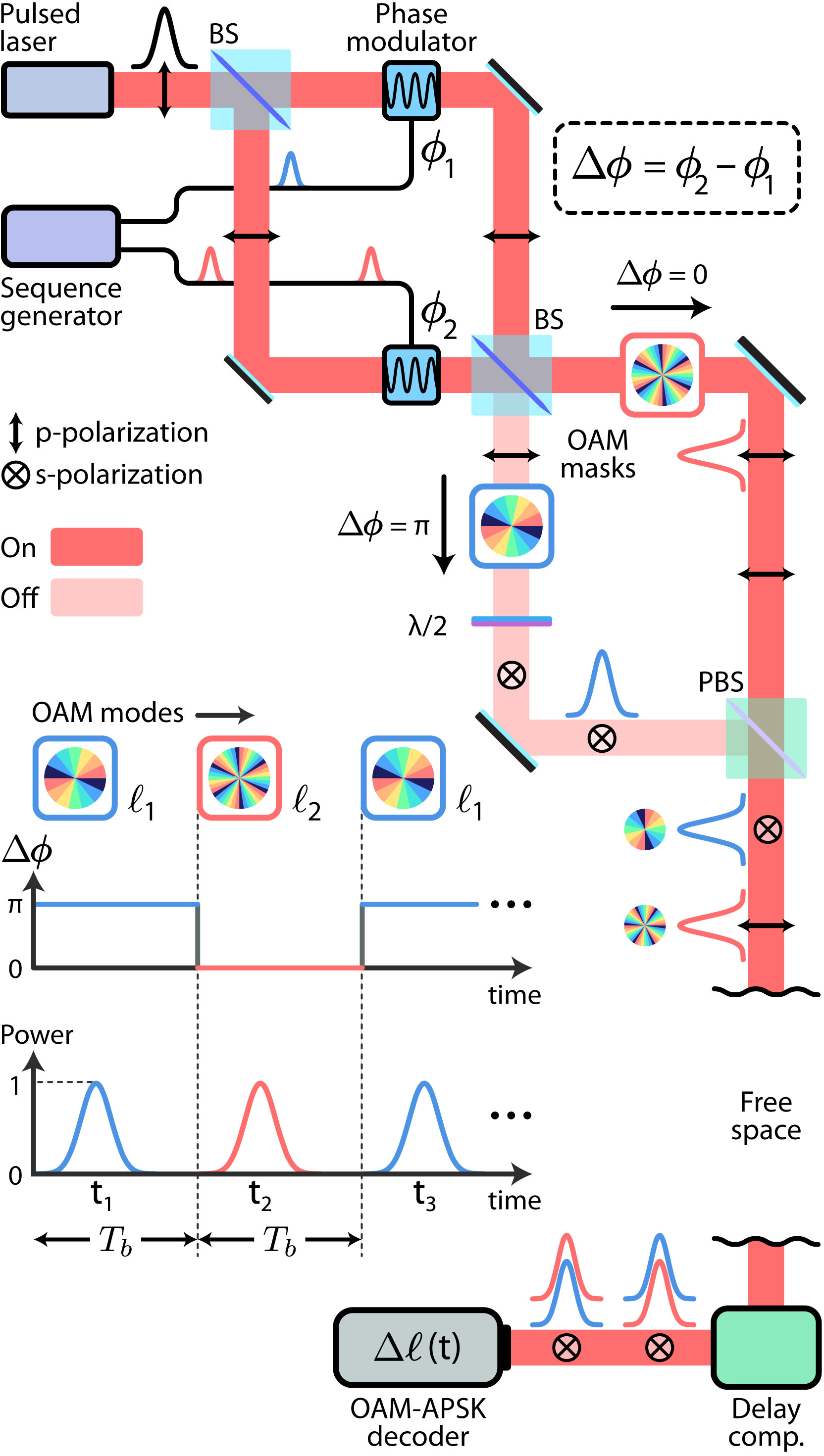}
		\caption[]{OAM-APSK phase-modulated scheme. Unlike the amplitude-modulated system shown in Fig.~1 of the main paper, this setup uses 100\% of the laser power. Here the sequence generator controls the relative phase-shift between photons in the two arms of the first Mach-Zehnder interferometer, and depending on this shift, which oscillates between 0 and $\pi$, light goes into the arm that imposes an OAM mode of either $\ell_1$ or $\ell_2$, respectively. BS: beam splitter; PBS: polarization beam splitter.}
		\label{fig:oamphmod}
	\end{figure}
	
	\begin{figure*}
		[!ht]
		\centering
		\includegraphics[width=1\linewidth]{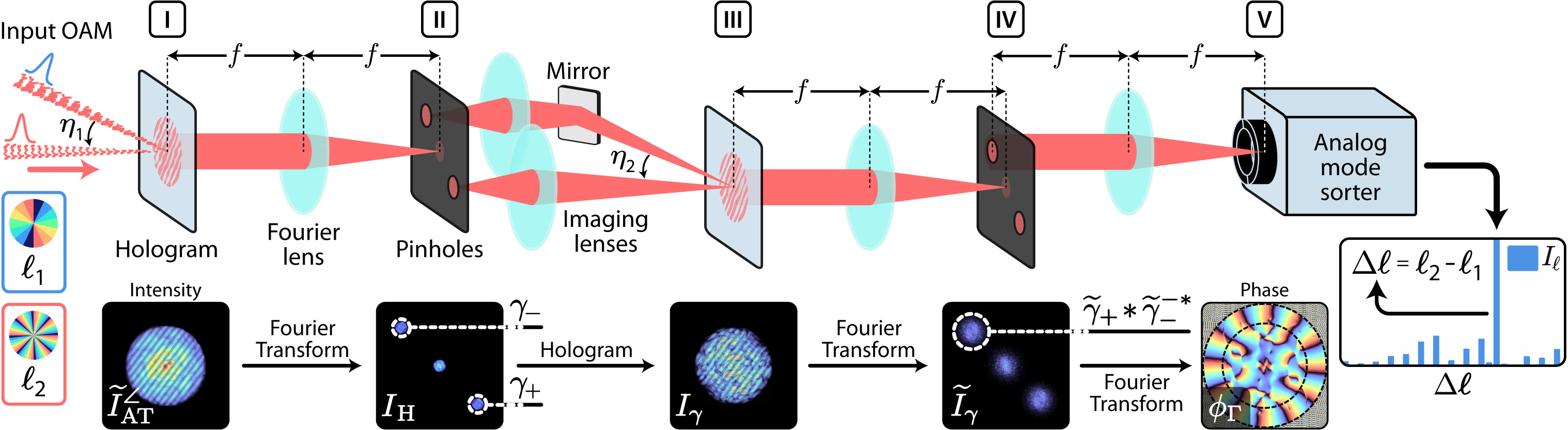}
		\caption[]{Physical implementation of OAM-APSK analog holographic mode sorting (AHMS). Different steps in the algorithm and physical setup are denoted by the column numbers I through V. We assume that it is possible to easily record and read-out physical holograms (gray screens) using something like a diffusive screen or nonlinear material. I) The first hologram is recorded, as in regular DHMS, of the interference between two OAM modes. II) We take a Fourier transform of the hologram using a lens, and at the Fourier plane we sample the spectrum $I_h$ using two pinholes on the spots that correspond to $\gamma_-$ and $\gamma_+$. III) The latter two fields are imaged to record another hologram, $I_{\gamma}$. IV) We use a lens to project the spectrum $\widetilde{I}_{\gamma}$ onto a pinhole screen again, this time only filtering out the field $\widetilde{\gamma}_+ \ast \widetilde{\gamma}_-^{-*}$. V) A final lens takes the latter's Fourier transform, yielding $\gamma_+\gamma_-^*$, from which we can extract $\phi_{\Gamma}$ and use an analog mode sorter (e.g.~a wavefront sensor or cascaded spatial light modulators) to measure the OAM-APSK mode $\Delta\ell$.}
		\label{fig:oamphys}
	\end{figure*}
	
	\section{Experiment}
	
	\subsection{Setup details}
	
	In this section we include a more detailed description of our experimental setup.
	
	On the source or encoding side, we used a He-Ne laser (Meredith Instruments 633nm Red HeNe Head) with a nominal power of 5 mW, and all optics were anti-reflection coated in the visible band. For encoding OAM modes, we opted for using a vortex phase plate (VPP) (RPC Photonics VPP-m633) with multiple 10 mm diameter OAM phase patterns lithographically printed on it. The VPP was placed in one arm of a Mach-Zehnder interferometer before the second beam-splitter and the other arm was empty (although in general it can also contain a separate VPP as shown in Fig.~6 of the main paper). 
	
	We verified the integrity of the OAM modes by interfering the  positively charged ($+\ell$) azimuthal charge modes with negatively charged ones ($-\ell$), which were created by using a Dove prism in one arm of a Mach-Zehnder interferometer (MZI) and by placing the VPP before the initial beam is split by the MZI. The intensity profiles of the OAM beams are shown in Fig.~\ref{fig:oamex}, and their symmetric annular shapes and flower-petal interference patterns verify the integrity of our VPP.
	
	In our experiment aimed to demonstrate a \textit{proof-of-concept} that DHMS can reduce phase noise in OAM-APSK signals. Therefore, we did not decide to use an exact AT phase screen implemented via spatial light modulator (SLM) or lithographically fabricated phase mask. More details on the mask are included in the next subsection.
	
	On the detector side we used a polarization beam splitter to separate the two beams by an angle of 1.9 degrees with respect to the normal axis of the detector. This corresponds to a distance of 52.5 mm$^{-1}$ from the origin in frequency space, or about half of the maximum bandwidth-limit. The CCD (iDS UI-2240-SE-C-HQ) we used has 1280 $\times$ 1024 pixels with a 4.65 $\mu$m pixel pitch, and we worked in-low exposure low-gain settings by attenuating the laser with a polarizer.
	
	Although we discard the polarization degree of freedom in our experiment, in theory we \textit{are} able to multiplex two different OAM modes that are on opposite sides relative to a central probe mode. For example, we can have one central OAM mode $\ell_{0} = 0$, a greater OAM mode $\ell_{+} = +2$, and a lesser OAM mode $\ell_{-} = -20$. We are limited in that having any more than three OAM modes in an individual hologram leads to serious OAM interference in the holographic phase measurement.
	
	\subsection{Turbulence strength characterization of phase screen}
	
	In order to characterize the phase-distorting mask used in OAM-DHMS experiments, we used two metrology techniques: phase-shifting interferometry and holographic phase measurement. We then calculate the phase structure functions from the holography results and compare them to theoretical estimates.
	
	Phase-shifting interferometry was done using a Zygo Verifire PE Fizeau interferometer, and the data was analyzed using the Zygo MetroPro software. We display both the intensity and phase measurement results in Fig.~\ref{fig:verifire}. It is clear from the interferogram in Fig.~\ref{fig:verifire}(a) that there are several fine-scale distortions relative to the scale of our beam diameter, which was about 2.5 cm wide. However, due to the significant variations, inhomogeneity, and imperfect flatness of the mask, the phase measurement in Fig.~\ref{fig:verifire}(b) had several fall-off errors, which resulted in gaps of the 2D phase surface. Indeed, the Verifire is suited to measuring an RMS phase variation of up to 15 waves, but we recorded a potential RMS of 52 waves (at 633 nm). Again, we emphasize that our experimental goal was to provide a proof-of-concept that OAM-DHMS could work even for extreme turbulence conditions, which we can confirm from the Verifire analysis. Although noise in the phase mask may not necessarily be precisely described by Kolmogorov statistics, the distortion is still significant enough that it corroborates the robustness of OAM-DHMS.
	
	Additionally, we provide a comparison of digital holographic phase measurements of an undistorted and clean Gaussian beam before inserting the phase-distorting mask shown in Fig.~\ref{fig:verifire}. This is performed with a setup similar to the DHMS scheme used in the main experiment, the main difference being that the phase-distorting mask is only inserted in one arm of the DHMs interferometer. In Fig.~\ref{fig:dhmsdr0}(a, b), we clearly see the impact on the original beam intensity, where the central lobe is fractured and a large fraction of power is sent into the side-lobes. The phase in Fig.~\ref{fig:dhmsdr0}(c, d) also shows clear distortion, where an originally uniform and circularly symmetric distribution is shifted off-center and there are additional low and high-frequency inhomogeneities.
	
	We give a more analytical description of the phase-distortion by calculating the phase structure function $D_{\phi}$ (versus separation distance $|r|$) of both the undistorted and distorted data, and these are compared with analytical estimates of the von-Karman structure functions $D_{\phi}^{vK}$ (Eq.~\ref{eq:atsfxn}) for different values of $D/r_0$. These results are shown in Fig.~\ref{fig:dhmsdr0-sfxn}, where there is a clear difference in magnitude and shape between the distorted (red) and undistorted (blue) structure functions. An intuitive way of looking at these results is by understanding that larger values of $D_{\phi}$ correspond to a smaller average correlation between points on the phase surface separated by a specific distance $|r|$. Therefore, the undistorted blue curve describes a more correlated (smaller magnitude) phase surface compared to the red curve (larger magnitude) that corresponds to the noisier distorted data. This is also evident from the fitted von-Karman structure functions, where we only need $D_{\phi}^{vK}$ corresponding to $D/r_0 \sim 7$ to encompass the undistorted $D_{\phi}$, but for the distorted data we need $D_{\phi}^{vK}$ corresponding to $D/r_0 \sim 27$. We note that the functional difference between $D_{\phi}$ (from data) and $D_{\phi}^{vK}$ is most likely due to an inadequate amount of low-frequency noise in the phase-distorting mask. This phenomenon can be seen from the way in which subharmonics improve the simulated AT phase structure functions to more closely match the theoretical $D_{\phi}^{vK}$ (Fig.~\ref{fig:atstruct}).
	
	In conclusion, although our characterization of the phase-distorting mask is not precise due to our lack of a high-quality wavefront imaging device, by using phase-shifting interferometry and simple digital holographic phase measurements, we have shown that the distortion is significant enough that it is a valid test for demonstrating the robustness of OAM-DHMS against turbulent noise.
	
	\subsection{Additional data}
	
	In Figure \ref{fig:oamres} we include additional experimental results similar to those shown in the main article, but these use different OAM mode pairs ($\ell_1, \ell_2$). As in the (0, 8) case shown in the main article, the cases included here show that DHMS signals (blue bars) are robust to both phase and amplitude distortion. The increase in SNR of DHMS results is not significant, even when our phase-distorting mask completely distorts the amplitude profiles of the OAM beams and their spectra. Further, the peak amplitude in DHMS results is unchanged for both the (0, 3) and (0, 5) cases.
	
	\section{OAM-APSK Extensions}
	
	\subsection{Phase-modulated OAM-APSK}
	
	In the main paper we introduced the OAM-APSK concept using an amplitude-modulation scheme, where we would switch between light pulses with one of two OAM modes using a Mach-Zehnder interferometer-style setup. However, knowing that the amplitude-modulated scheme would lead to power-loss, in Fig.~\ref{fig:oamphmod} we introduce a phase-modulated OAM-APSK scheme. In essence, the electro-optic modulated gates originally in both arms of the MZI are replaced with phase modulators. Using two of the latter, we can ensure higher switching accuracy.
	
	The phase difference $\Delta\phi = \phi_2 - \phi_1$ between the two MZI arms determines which OAM path is taken by the input light, e.g.~$\Delta\phi = 0$ or $\pi$ corresponds to $\ell_2$ and $\ell_1$, respectively. Further, in order to conserve 100\% of the power, we use a half-wave plate and a polarization beam splitter to recombine the separate OAM mode paths in the second MZI setup. After propagation, delay compensation, and polarization recombination, digital sorting of APSK modes proceeds as usual at the decoder.
	
	\subsection{OAM-APSK analog holographic mode sorting}
	
	Currently the most practical way of performing holographic sorting is through digital means, mostly due to the low accessibility to rewritable holographic materials with fast response rates. However, here we describe a general layout for a possible analog holographic mode sorting scheme (AHMS) for OAM-APSK.
	
	Figure \ref{fig:oamphys} shows a linear AHMS scheme that makes use of only lenses, holographic recording materials, and pinhole screens, with a final analog mode sorter at the end. In essence, we must transform the input field of OAM modes into the electric field $\gamma_+\gamma_-^*$ such that the analog mode sorter positioned at the end can operate on the phase $\phi_{\Gamma}$. 
	
	To achieve the latter, we can use lenses to project the angular spectrum $I_{\rm H}$ from the recorded hologram $\widetilde{I}_{\rm AT}^{\angle}$ onto other propagation planes for analog optical information processing [Fig.~\ref{fig:oamphys}(I)]. In this way, we can extract the two fields $\gamma_+$ and $\gamma_-$ by using a simple pinhole mask at the Fourier plane [Fig.~\ref{fig:oamphys}(II)]. 
	
	For the next step in AHMS (analogous to DHMS) where we must multiply $\gamma_+$ by $\gamma_-^*$, there is no need to perform overly-involved physical optical conjugation if we can obtain $\gamma_-^*$ by using another hologram. In order to achieve this, we simply image the $\gamma_-$ and $\gamma_+$ fields onto another holographic material plane [Fig.~\ref{fig:oamphys}(III)], tilting one field at an angle $\eta$ (where $k_{\eta} = k\sin\eta$):
	\begin{equation}
	I_{\rm H}^{\angle} = \gamma_+ + \gamma_- e^{ik_{\eta}(x + y)}.
	\end{equation} The intensity of the latter yields our second hologram $I_{\gamma}$:
	\begin{equation}
	I_{\gamma} = |I_{\rm H}^{\angle}|^2 = |\gamma_+|^2 + |\gamma_-|^2 + \gamma_+\gamma_-^{*}e^{ik_{\eta}(x + y)} + \rm c.c.
	\end{equation}
	
	Now the field $\overline{\gamma}$ that we want is physically recorded, but still inaccessible and enmeshed with the DC noise. However, if we take another Fourier transform [Fig.~\ref{fig:oamphys}(III, IV)], we can physically separate the $\overline{\gamma}$ spectral components in another Fourier plane as the field $\widetilde{I}_{\gamma}$:
	\begin{equation}
	\widetilde{I}_{\gamma} = \widetilde{\gamma}_{0} + (\widetilde{\gamma}_{+} \ast \widetilde{\gamma}_{-}^{-*})\ast\delta(\mathbf{k} + k_{\eta}) + \rm c.c.
	\end{equation}
	Here we denote the DC components by $\widetilde{\gamma}_0$ and are mainly focused on the offset interference components.
	
	The final AHMS step is straightforward, where we only require one of the offset spectral terms. We filter the field with another pinhole screen and perform the Fourier transform on one of either offset fields in $\widetilde{I}_{\gamma}$. This turns the convolution $\widetilde{\gamma}_{+} \ast \widetilde{\gamma}_{-}^{-*}$ into the desired electric field $\gamma_+\gamma_-^*$ containing $\phi_{\Gamma}$ [Fig.~\ref{fig:oamphys}(V)]:
	\begin{equation}
	\overline{\gamma} = \gamma_{+}\gamma_-^* = \mathcal{F}\{\widetilde{\gamma}_+ \ast \widetilde{\gamma}_-^{-*} \}.
	\end{equation}
	
	\begin{figure*}
		\centering
		\includegraphics[width=1\linewidth]{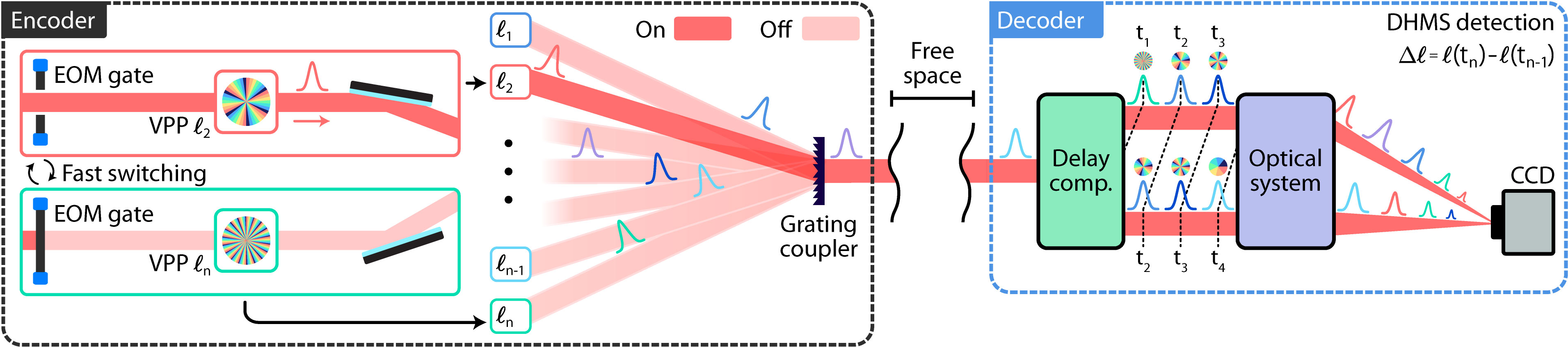}
		\caption[]{Conceptual illustration of a scheme for multi-valued OAM-APSK bit encoding and decoding. A laser is split into $n$ beams, each of which is incorporated with a phase mask to impose a certain OAM $\ell_j$ ($j=1,2,\cdots,n$) and a high-speed electro-optic gate (EOM gate) for fast switching to produce optical communication pulses. The EOM gates are controlled such that only one gate is open during a bit time slot to produce an optical communication pulse carrying a certain OAM. the OAM beams are combined, say, via a grating coupler. At the decoder side, the scheme from Fig.~\ref{fig:2} can be used for DHMS detection of the multi-valued OAM APSK signals. }
		\label{fig:oamasmvdpsk}
	\end{figure*}
	
	\begin{figure*}
		\centering
		\includegraphics[width=0.8\linewidth]{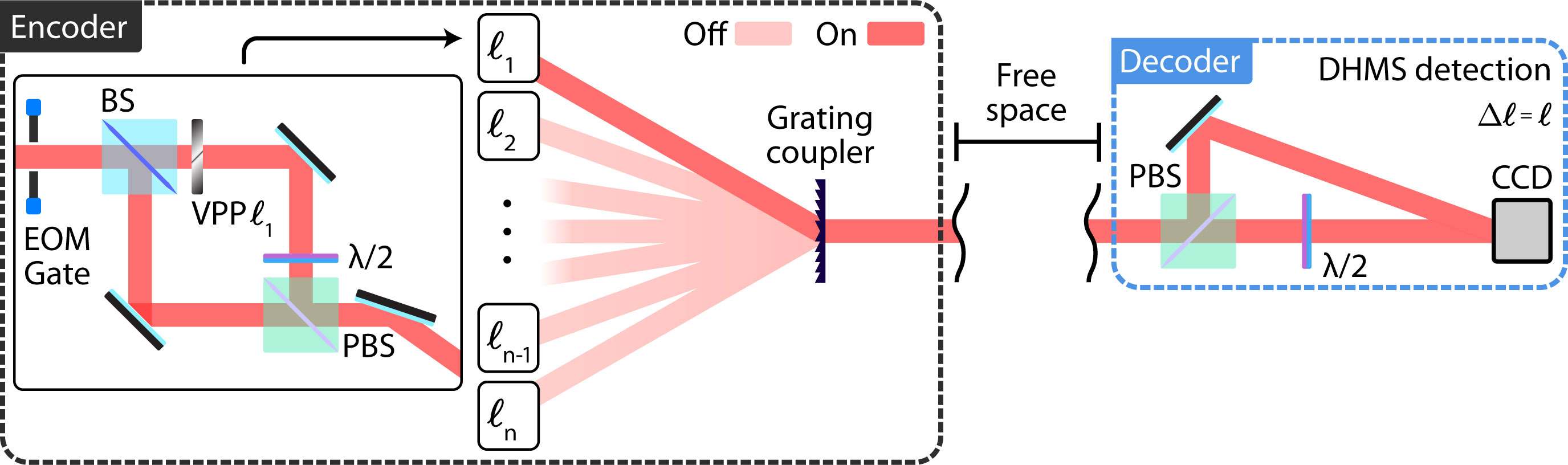}
		\caption[]{OAM-APSK delay-less coherent multiplexing communication link. This scheme is similar to the link shown in the main article, but DHMS detects an amplitude-dependent OAM mode difference $\Delta\ell$ instead of a purely phase-dependent one. Here each individual output before the coupler ($\ell_1, \ell_2, \dots,\ell_{n-1}, \ell_n$) already contains two beams that are orthogonally polarized and combined, therefore already containing an independent value of $\Delta\ell$. Since one arm in each Mach-Zehnder interferometer lacks an OAM phase mask or VPP, $\Delta\ell = \ell - 0 = \ell$ and the outputs are simply denoted by $\ell$. As in the original APSK scheme that uses delayed pulses, we use an electro-optic modulator gate to modulate the laser amplitude and switch between $\ell$ outputs. DHMS is used to decode the OAM-APSK signal by using a CCD camera at the output. BS: beam splitter; PBS: polarization beam splitter.}
		\label{fig:oamascoh}
	\end{figure*}
	
	\begin{figure*}
		\centering
		\includegraphics[width=1\linewidth]{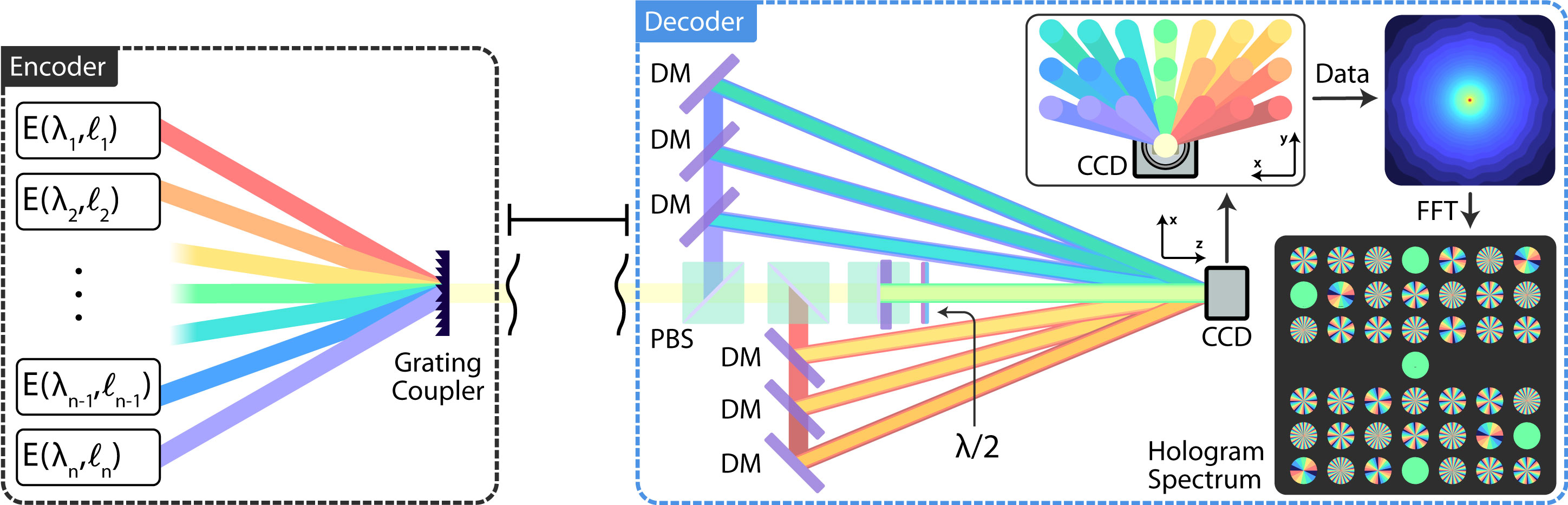}
		\caption[]{OAM-APSK broadband angular multiplexing communication link. OAM beams with different $\ell$ and $\lambda$ (from independent fast electro-optic modulator-switched encoders) are multiplexed via a grating coupler. In the decoding side, several narrow-band polarizing beam splitters are used to separate different wavelength components from the incoming beam. These beams are then directed towards the CCD camera using \textit{very} narrow-band dichroic mirrors (DM), with each wavelength corresponding to a different angle of incidence on the CCD. After taking a fast-Fourier transform (FFT), we see that the wavelength-angle correspondence results in a spatial frequency grid where the wavelengths are coupled to separate OAM spectral distributions which can be independently mode-sorted. PBS: polarization beam splitter.}
		\label{fig:oamasbrdam}
	\end{figure*}
	
	\begin{figure*}
		\centering
		\includegraphics[width=1\linewidth]{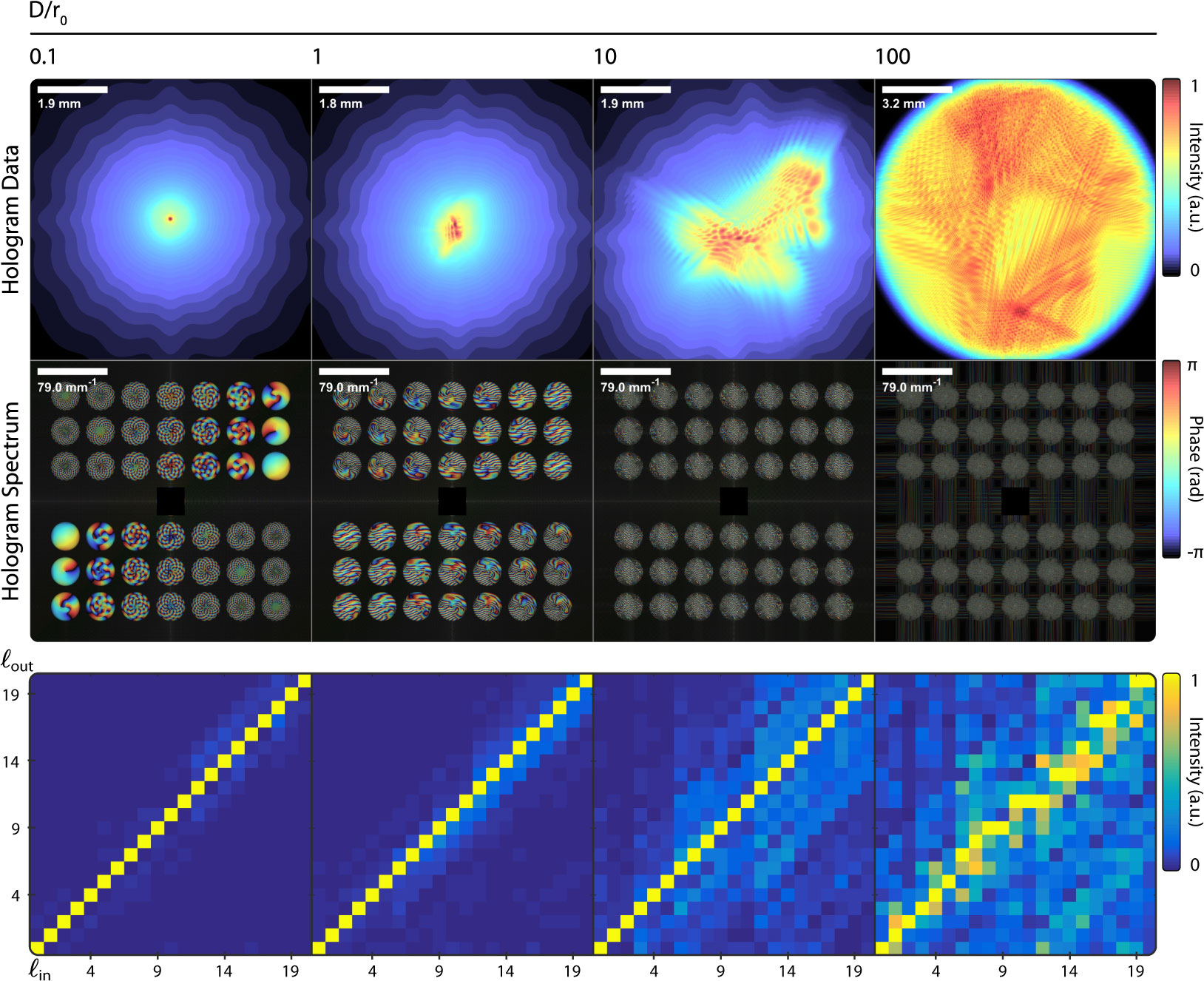}
		\caption[]{OAM crosstalk simulations of OAM-APSK broadband angular multiplexing communication link. Columns represent results for four values of $D/r_0$ (0.1, 1, 10, and 100). The first, second, and third rows show the hologram data, angular spectra, and the APSK-DHMS crosstalk matrices, respectively. In the second row, the separate circular areas correspond to different OAM modes and wavelengths. The last row shows crosstalk matrices plotted as the DHMS sorting amplitudes of output OAM modes $\ell_{\rm out}$ for corresponding input OAM modes $\ell_{\rm in}$.}
		\label{fig:oamasbrdamxm}
	\end{figure*}
	
	With our final result, we have a few options for analog sorting of OAM modes. The most straightforward method would be conventional mode sorting, where we multiply the electric field $\gamma_+\gamma_-^*$ by OAM basis modes produced using either spatial light modulators (SLM) or static phase masks. This is esentially the fastest method, since the output of SLM sorting channels can be recorded with fast photodetectors. However, although $\phi_{\Gamma}$ contains a clean OAM profile, we have also preserved the still potentially distorted amplitude, which might lower the SNR of OAM-APSK signals obtained from mode sorting.
	
	Alternatively, we can opt for a wavefront sensor as our AHMS decoder. In this way, we can exclusively measure $\phi_{\Gamma}$ while ignoring the field amplitude. We would still have to sort modes digitally via integration of OAM basis state inner products as in DHMS, but the signals would be much clearer. The trade-off would be a reduction in speed, but we are still saving ourselves the step of having to calculate any FFTs.
	
	\subsection{Additional OAM-APSK systems}
	
	Here we describe several alternative OAM-APSK frameworks, including one coherent delay-less scheme and two broadband OAM-APSK schemes.
	
	Using the same system described in the main paper, we can multiplex several pulse trains for OAM-APSK by using a grating coupler as shown in Fig.~\ref{fig:oamasmvdpsk}. Each arm before the grating corresponds to a unique OAM mode input, although we still require two OAM signals in order to determine a single OAM-APSK bit at the decoder side. This is essentially a multi-valued OAM scheme, and due to the requirement that consecutive pulses must interfere, this scheme requires that all OAM signals in each separate arm are mutually coherent.
	
	An alternative coherent link is shown in Fig.~\ref{fig:oamascoh}. Compared to the original scheme in Fig.~1 of the main paper, in this so-called delay-less scheme all OAM-APSK encoding is \textit{amplitude} based. We state this in the sense that each OAM-APSK signal being combined by the grating coupler is independent, since two orthogonally polarized beams are separately generated by independent sources before the coupler. Each source contains a different OAM-APSK mode $\Delta\ell = \ell$, where this latter equivalence is due to that one beam in the independent interferometers has an OAM mode of $\ell = 0$. It may be beneficial to use this scheme if it proves difficult to implement a delay compensation system as shown in the main paper. However, the trade-off here is that each arm requires its own interferometer set-up, meaning more components and aligning are necessary. For the simplest case where there are only two possible modes being multiplexed, e.~g.~$\ell = \{0, 1\}$, the original APSK scheme allows for three unique bits: $\Delta\ell = \{0,+1,-1\}$. However, the delay-less scheme only allows two bits, $\Delta\ell = \{0,+1\}$, since a negative OAM difference (from consecutive bits/pulses) cannot be distinguished. 
	
	Although coherent OAM-APSK schemes are limiting in terms of multiplexing, using the optical wavelength as an additional degree of freedom significantly increases the multiplexing flexibility. However, in order to further increase the information density of a broadband OAM-APSK link, we can make full use of the hologram spectrum. Fig.~\ref{fig:oamasbrdam} describes what we term a broadband OAM-APSK angular multiplexing link, where we sacrifice power efficiency (losses due to diffraction coupling) and resolution (partitioning frequency space for multiple OAM beams) for space and speed. In Fig.~\ref{fig:oamasbrdamxm} we show simulation results for this angular multiplexing link, where each circular area in the hologram spectrum contains a distinct azimuthal mode that ranges from 0 to 20. For moderate strength turbulence with $D/r_0 = 10$, a negligible amount of crosstalk is evident in the matrix, even though the amplitude and phase profiles are indistinct from the weak turbulence case of $D/r_0 = 0.1$. Crosstalk increases dramatically with $D/r_0 \sim 100$, and we believe that the main reason for this is the vignetting of beams by the digital aperture. This was not a problem for the results in Fig.~4 of the main paper since the beam was never clipped, but here we must sacrifice a larger aperture at our detector in order to have smaller and more numerous partitions in the angular spectrum. Indeed, Laguerre-Gaussian beams have the property that larger values of OAM lead to radically faster divergence during propagation. Thus in order to resolve this issue, we must figure out a way to ensure that all OAM modes have the same diameter when they are incident on the detector.
	
	In short, here we demonstrated that it is possible to computationally multiplex 21 independent and incoherent OAM signals in one hologram. However the resolution of DHMS is limited by the array size used for the hologram FFT, and the pixel size of our detector limits the maximum spatial frequency. In these angular multiplexing scenarios there is inevitably a trade-off between number of channels and increased resolution or availability of larger azimuthal mode values in a single channel.

\end{document}